\documentclass{aa}
\usepackage[varg]{txfonts}
\usepackage[utf8]{inputenc}
\usepackage{graphicx}
\usepackage{array}
\usepackage{hyperref}
\usepackage{multirow}
\usepackage{natbib}
\usepackage{enumitem}
\usepackage{pifont}
\usepackage{xfrac}
\usepackage{nicefrac}
\usepackage{mathtools}
\usepackage{amsmath}

\DeclarePairedDelimiter\floor{\lfloor}{\rfloor}

\usepackage{ulem}

\hypersetup{
    colorlinks=true,
    linkcolor=blue,
    citecolor=blue
}

\begin{document}

\title{A stochastic and analytical model of hierarchical fragmentation}
\subtitle{The fragmentation of gas structures into young stellar objects in the interstellar medium}

\author{Thomasson B.$^{1}$
\and Joncour I.$^{1}$
\and Moraux E.$^{1}$
\and Motte F.$^{1}$
\and Louvet F.$^{1}$
\and González M.$^{2}$
\and Nony T.$^{3}$}

\institute{
$^{1}$ Univ. Grenoble Alpes, CNRS, IPAG, 38000 Grenoble, France\\
$^{2}$ Universidad Internacional de Valencia (VIU), C/Pintor Sorolla 21, E-46002 Valencia, Spain\\
$^{3}$ INAF - Osservatorio Astrofisico di Arcetri, Largo E. Fermi 5, 50125 Firenze, Italy}

\date{Received <date> /
Accepted <date>}

\abstract {
Molecular clouds are the most important incubators of young stars clustered in various stellar structures whose spatial extension can vary from a few AU to several thousand  AU. Although the reality of these stellar systems has been established, the physical origin of their multiplicity remains an open  question.}{Our aim was to characterise these stellar groups at the onset of their formation by quantifying both the number of stars they contain and their mass using a hierarchical fragmentation model of the natal molecular cloud.} {
We developed a stochastic and predictive model that reconciles the continuous multi-scale structure of a fragmenting molecular cloud with the discrete nature of the stars that are the products of this fragmentation. In this model a gas structure is defined as a multi-scale object associated with a subregion of a cloud. Such a structure undergoes quasi-static subfragmentation until star formation. This model was implemented within a gravo-turbulent fragmentation framework to  analytically follow the fragmentation properties along spatial scales using an isothermal and adiabatic equations of state (EOSs).
}{We highlighted three fragmentation modes depending on the amount of fragments produced by a collapsing gas structure, namely a hierarchical mode, a monolithic mode, and a {mass dispersal} mode. Using an adiabatic EOS we determined a characteristic spatial scale where further fragmentation is prevented, around a few tens of AU. We show that fragmentation is a self-regulated process as fragments tend to become marginally unstable following a $M \propto R$ Bonnor--Ebert-like mass--size profile. Supersonic turbulent fragmentation structures the cloud down to $R \approx 0.1$~pc, and gradually turns into a less productive Jeans-type fragmentation under subsonic conditions {so hierarchical fragmentation is a scale dependant process}.} {Our work suggests that pre-stellar objects resulting from gas fragmentation, have to progressively increase their accretion rate in order to form stars. A hierarchical fragmentation scenario is compatible with both the multiplicity of stellar systems identified in Taurus and the multi-scale structure extracted within NGC~2264 molecular cloud. This work suggests that hierarchical fragmentation is one of the main mechanisms explaining the presence of primordial structures of stellar clusters in molecular clouds.}

\titlerunning{A stochastic and analytical model of hierarchical fragmentation}

\maketitle

%
%

\section{Introduction}
\label{Section:Introduction}

Molecular clouds are the privileged environment for the formation of young stars. These stars are rarely born in isolation \citep{LadaLada2003} as they emerge in clusters or pair together in multiple systems that can extend from a few AU to several hundred \citep{duchene_stellar_2013} or thousand  AU \citep{joncour_multiplicity_2017}. In addition,
the newly born stars are characterised by an initial mass function (IMF) that represents the probability density function (PDF) of their mass distribution. Among star formation processes operating in molecular clouds, some models favour local stochastic processes to regulate (i) the mass of the stars, (ii) their spatial distribution, and (iii) the multiplicity of stellar groups through dynamical interactions between protostars themselves, for example coalescence \citep{Dib2017}; dynamical ejection or capture \citep{reipurth2014}; stellar outflows \citep{Adams1996}; or competitive accretion \citep{bonnell2001}.
These local stochastic processes at small scale would erase residual imprints of large-scale physical processes structuring both the cloud and the pre-stellar cores, regarding the IMF and multiplicity of stellar systems.

Alternatively, it has been proposed that the stellar IMF may be directly  inherited from the mass distribution of the parental dense gas cores at a larger scale \citep{Motte1998, Sadavoy2010, offner2014}. The increasing number of observations at different spatial resolutions and the improved performance of recent instruments have revealed that these dense structures of gas are actually made up of several buried substructures \citep{offner2012, pokhrel_hierarchical_2018, thomasson2022}. The existence of such substructures is often interpreted as the result of core fragmentation, although their physical properties may depend on the instrumental resolution \citep{Louvet2021}, which complicates their full characterisation.

Consequently, the fragmentation of molecular clouds all the way down to dense cores that may in turn subfragment to constitute the birth sites of young stellar systems is one of the driving processes shaping both the IMF and young stellar groups. In that case, the spatial distribution of young stars would simply mimic the structure of the cloud. To support this idea, multiplicity analysis of ultra-wide stellar systems in Taurus \citep{joncour_multiplicity_2017} suggests a fragmentation cascade scenario for their formation. Further studies unravel the presence of dense stellar  Nested Elementary STructures (NESTs) in star-forming regions \citep{joncour2018nests, Gonzalez2021} whose origins may be attributed to the cloud fragmentation into gas cores clusters that in turn may fragment further into young stars. Hence, these stellar groups are suspected to be the pristine vestiges of star formation, and the emergence of these groups may be the result of physical mechanisms that  regulate both the hierarchical fragmentation and the molecular cloud's structure at larger scale. Hereafter we refer to a local concentration of stars, gravitationally bound or not, within a cloud as a stellar structure or stellar system.

{
Structures of different morphologies such as filaments, gas clumps, or pre-stellar cores may emerge at different spatial scales inside molecular clouds depending on the relevant physical processes shaping these scales (e.g. turbulence, magnetic field, rotation, or even stellar feedback). In that context, molecular clouds are characterised by their multi-scale nature. To investigate the origin of stellar systems through the fragmentation of the ISM, every process that influences the structure of the cloud at their relevant scale needs to be accounted for. In this work we focus on the fragmentation of gas clumps considering turbulence and the thermodynamics of gravitational collapse.}


Quasi-static gravitational instabilities \citep{jeans1902} constitute the basis of our current understanding concerning the local collapse of self-gravitating gas clumps into young stars. Throughout their evolution, molecular clouds condense into self-gravitating structures from which stars eventually form. A self-gravitating clump may collapse monolithically \citep{larson1969_fcore, shu_self_similar_1977} under isothermal pressure free condition \citep{tohline1980}, but may also fragment in order to form multiple dense structures \citep{hoyle1953, Guszejnov_isothermalfrag, vazquez-semadeni_global_2019}. We explore in this work the possibility that fragmentation of molecular clouds shapes the stellar properties directly from cloud physical properties.

Provided that a self-gravitating clump neither loses mass during its evolution nor accretes mass from its neighbouring environment, the number of Jeans masses contained in the initial structure increases as it contracts quasi-statically \citep{hoyle1953, hunter1962}. New gravitationally unstable regions then appear within the initial structure, causing its subfragmentation and so on \citep{guszejnv2018_scalefree, vazquez-semadeni_global_2019}. The collapse of each fragment may eventually end as the gas structure becomes opaque to its own radiation and stabilises. This stage is known as the first Larson core \citep{hoyle1953, larson1969_fcore}, and the following adiabatic phase prevents any other fragmentation events. However, to achieve total stabilisation, the structure should exert an effective pressure force large enough to support its own mass. Thus, the final mass and size properties of these structures should depend on the growth rate of the instability during collapse as mass can be dynamically added or removed, but also on the degree of instability of the initial structure \citep{tohline1980}. 

Hierarchical fragmentation of clumps within other clumps has been analytically described by \cite{hoyle1953} using geometric sequences tracing the evolution of the mass of the fragments after series of fragmentation events. Since then, stochastic models of fragmentation were developed {for solely to derive the stellar IMF} \citep{larson1973_fragmentation, elmegreen1983} or to study the transfer of the angular momentum throughout many fragmentation events \citep{zinnecker1984}. {
{Even though it has not been done,} these {discrete and} stochastic approaches are adapted to investigate the multiplicity of the resulting stellar groups as it is possible to simply count the number of stars at the end of the process. However, these models were not designed to account for the physical mechanisms  or the thermodynamics that structure the parental cloud from which young stars emerge.}

On the other hand, more recent analytical approaches based on the interplay between supersonic turbulence and gravity in the interstellar medium {account for the physical processes shaping the continuous multi-scale structure of molecular clouds} \citep{H&C2008, hopkins2012_excursion, GuszejnovHopkins2015}. In these gravo-turbulent models, self-gravitating structures are formed locally within the turbulent cloud through small density enhancements that in turn collapse or eventually subfragment. 
{In this framework, the fragmentation and collapse of a structure is} modulated by the balance between (i) gravity that triggers local collapse within a structure, (ii) turbulence {that supports} the collapse at large scale, but also induce local density enhancements that may collapse, and (iii) thermal pressure. {Although these gravo-turbulent models provide a continuous description of a cloud, they do not consider the discrete nature of fragmentation so the multiplicity of stellar groups remains unknown.}


Our aim is then to connect the continuous and diffuse structure of the cloud that regulates the fragmentation processes with the result of this fragmentation (i.e. a collection of countable and discrete distribution of stars). 
{Based on a combined stochastic and gravo-turbulent approach, we introduce a framework to model hierarchical fragmentation and monitor the multi-scale structure of a cloud shaped by hierarchical fragmentation. Implementing a set of physical processes at different scales, this model allows us to predict the characteristic mass of last fragmenting pre-stellar structures as well as the characteristic clustering of the pristine stellar groups produced through hierarchical fragmentation.}

The paper is organised as follows. In Sect. \ref{sec:section1} we establish a general stochastic model that describes the successive hierarchical subfragmentation of local regions within a cloud.  In Sect. \ref{sec:section2} we implement our model within the framework of the gravo-turbulent fragmentation, and assess the fragmentation properties. Then, in Sect. \ref{sec:section3} we constrain our model to existing observational data and evaluate its current limits {before concluding this work in Sect. \ref{sec:section4}.

%
%

\section{Description of the model}
\label{sec:section1}

The following stochastic model aims   to describe the multi-scale fragmentation of any extended region of a cloud and   to provide an analytical framework to investigate the outcome of cloud fragmentation in terms of fragments mass and multiplicity. In this model, a dense structure represents a multi-scale object whose substructures depend on the local physical processes that regulates its fragmentation until star formation.


\subsection{A stochastic and geometrical multi-scale model}
\label{Sec:A stochastic and geometrical multi-scale model}

\begin{figure*}
    \centering
    \includegraphics[width=16cm]{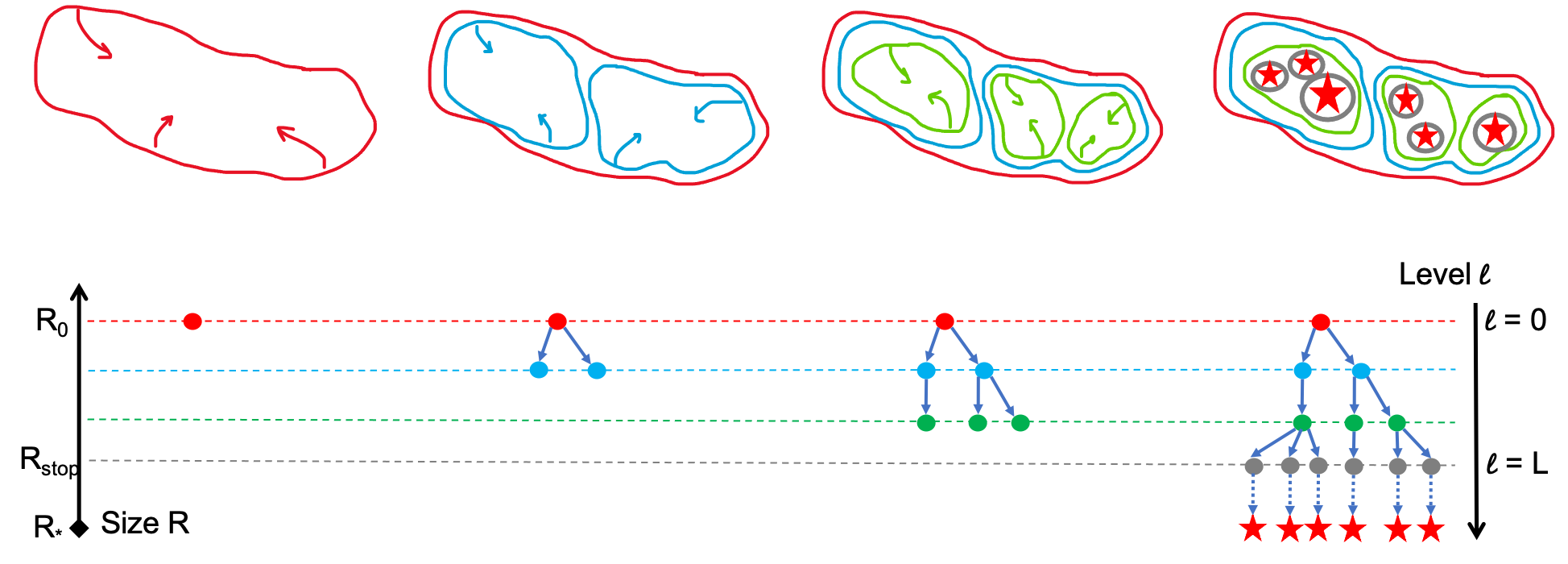}
    \caption{Schematics of the hierarchical fragmentation process of a cloud with its network representation. Each spatial scale $R$ is associated with a discrete level of fragmentation $l$. The fragmentation cascades from large to small scales, down to a level $l = L$ corresponding to a scale $R_\text{stop}$ beyond which we assume the objects no longer fragment, but collapse into a single star of scale $R_*$.}
    \label{fig:fragmentation_network}
\end{figure*}

To introduce our model of fragmentation, we define an extended structure as a subregion within the cloud that undergoes local gravitational collapse and may fragment into several parts with smaller spatial extent, called children. Then these children become the parents of the next generation and subfragment further. This operation is repeated for increasingly smaller scales (see Fig. \ref{fig:fragmentation_network}). The main challenge of such a model is to account for the discrete nature of fragmentation, which results in the formation of point-like and countable stars, but also on the continuous and multi-scale nature of the interstellar medium from which these stars form. Therefore, the model used to describe this cascade of fragmentation has to consider the duality between both the discrete product of fragmentation and the continuous origin of its media. Hence, we discretise the spatial dimension into several scales $R_l$, associated with the fragmentation level $l$. The level $l+1$ defines an additional fragmentation step at lower spatial scales (i.e. $R_l > R_{l+1}$), such that hierarchical fragmentation cascades from large fragments to smaller fragments. 
{The design of the following model ressembles the fragmentation model of \cite{larson1973_fragmentation} that stochastically described the mass hierarchy resulting from fragmentation. In this work we consider the spatial hierarchy resulting from fragmentation. This different approach allows us to model the multi-scale structure of the interstellar medium and implement relevant physical processes at specific spatial scales (see Sect. \ref{Sec:Theoretical framework}).}

Due to non-isothermal processes on small scales (e.g. the first Larson core, \citealt{larson1969_fcore}) the fragmentation cascade results in the formation of children that are eventually unable to subfragment but instead collapse to form individual stars. We define by $R_\text{stop}$ the scale beyond which no further fragmentation occurs. At this point, one fragment produces exactly one single young star. We choose to represent this process with a multi-layered network of connected nodes (Fig. \ref{fig:fragmentation_network}) in which each level $l$ is populated by fragments whose typical size corresponds to the spatial scale $R_l$ (Fig. \ref{fig:fragmentation_network}) while the genealogy between fragments is represented by directed edges from parents to children.

As we consider the geometrical extension of fragments, the maximum number of children of size $R_{l+1}$ that can be inserted inside a parent of size $R_{l}$ is limited by the filling volume of the parent. In fact, a parent of scale $R_l$ cannot contain too many children of size $R_{l+1}$ enclosed in its borders, unless these children overlap or merge together. To allow the existence of distinct children between two successive levels, a fragmentation event can only be triggered {at} a 
sufficiently high scaling ratio {$r$, where} $r = \dfrac{R_{l}}{R_{l+1}} > 1$. Assuming that the children do not intersect, the maximum number of children $n_{l, \text{max}}$ that can exist within a parent in three dimensions, without prior about the geometrical shape, is the 
ratio of the volume of the parent at scale $R_l$ to the volume of its children at scale $R_{l+1}$:

\begin{equation}
    n_{l, \text{max}} \sim r^{3}
    \label{eq:nmax_r}
.\end{equation}

Between two successive levels $l$ and $l+1$, the fragmentation outcome (i.e. the children masses and multiplicity) is regulated both by the number of produced children and by the fraction of the parental mass that each child inherits (Fig.\ref{fig:formation_procedure}). We define the discrete probability density function $p_l(n_l)$ as the probability that at a level $l$, a parent of mass $M_l$ fragments into $n_l$ children such that

\begin{equation}
    \displaystyle \sum_{n_l = 1}^{n_{l, \text{max}}} p_l(n_l) = 1
,\end{equation}

\noindent where $n_{l, \text{max}}$ is the maximum number of children possibly generated at the level $l+1$ given the geometrical constraints of Eq.\ref{eq:nmax_r}. The expected number of fragments $\overline{n_l}$ produced at the level $l+1$ by one parent at the level $l$ is given by

\begin{equation}
    \overline{n_l} = \displaystyle \sum_{n_l = 1}^{n_{l, \text{max}}} n_l \times p_l(n_l)
.\end{equation}
{The expected number of fragments $\overline{n_l}$ can be computed from any underlying probability distribution $p_l(n_l)$. As we aim to evaluate the general characteristics of fragmentation throughout the spatial scales, $\overline{n_l}$ is the only relevant parameter in the following.}

Next, a parent at level $l$ splits its mass $M_l$ between its $n_l$ children at level $l+1$. We can define the mass efficiency $\epsilon_{l}$ as the fraction of mass inherited by the children from their parent,

\begin{equation}
     \epsilon_{l} = \frac{\displaystyle \sum_{i=1}^{n_l} M_{l+1,i}}{M_l}
     \label{eq:epsilon_reservoir}
,\end{equation}

\noindent where the index $1 \leqslant i \leqslant n_l$ points to the $i$-th child produced. The mass reservoir $\epsilon_{l} M_{l}$ is then partitioned between the $n_l$ children with respect to a partition function $\psi_{l,i}$ describing the fraction of mass the $i$-th child inherits from the reservoir, such that

\begin{equation}
    \sum_{i=1}^{n_l} \psi_{l,i} = 1
    \label{eq:norm_omega}
.\end{equation} 

Consequently, the $i$-th child inherits a mass $M_{l+1,i} = \psi_{l,i} \epsilon_{l} M_{l}$ and we can verify that Eq.\ref{eq:epsilon_reservoir} holds (see Fig.\ref{fig:formation_procedure}). In this work we don't impose additional constraints on the partition function $\psi_{l,i}$. Next, we focus on the expected outcome of fragmentation by computing the average number of children produced at any scale $R_l$ (Sect.\ref{sec:Average number of fragments produced}) and their average mass (Sect.\ref{sec:Average mass of the fragments}). In particular we introduce two parameters that describe the continuous spatial variations in the number of fragments and their mass.

\begin{figure*}
    \centering
    \includegraphics[width=16cm]{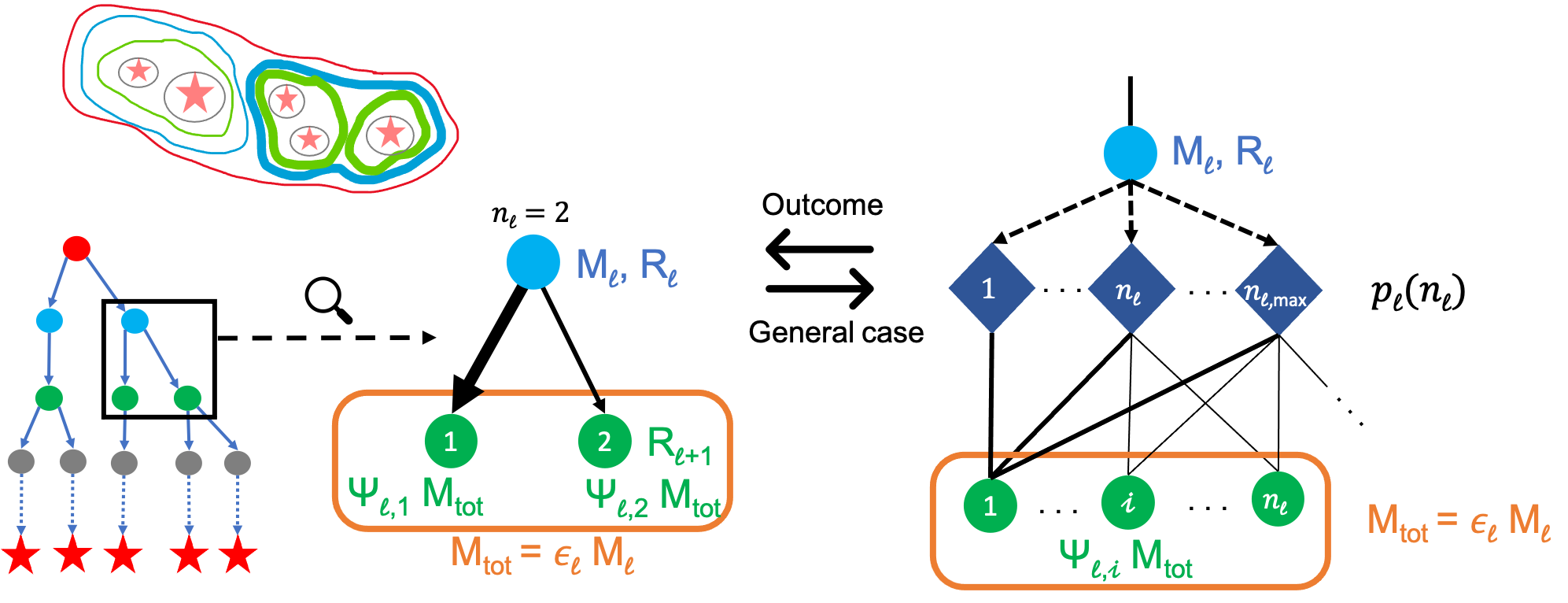}
    \caption{
    {Microscopic description of our hierarchical fragmentation model.} \textit{Left:} Fragmentation procedure of a single parent contained in a multi-scale structure of size $R_0$. Any parent at a level $l$, of scale $R_l$ and mass $M_l$, can be selected in order to evaluate the number $n_l$ of fragments it produces and then determine the fraction of its mass $\epsilon_l \psi_{l,i}$ distributed to the $i$-th child. The size of the arrow represents the amount of mass inherited by a child. \textit{Right:} Parent's probability $p_l(n_l)$ to produce $n_l$ children before transferring a proportion $\epsilon_l \psi_{l,i}$ of its mass to each of its children.}
    \label{fig:formation_procedure}
\end{figure*}


\subsection{Average number of fragments produced}
\label{sec:Average number of fragments produced}

\subsubsection{The fragmentation rate $\phi(R)$}
\label{sec:section1-1-1}

Considering a cloud of size $R_0$, at any level $l$ a parent has the probability $p_l(n_l)$ to fragment into $n_l$ children. Hence, all the parents $n_\text{par}$ on this level $l$ produce on average a total of $\langle N_\text{tot}(R_{l+1}) \rangle$ children, corresponding to

\begin{equation}
    \langle N_\text{tot}(R_{l+1}) \rangle = \displaystyle \sum_{n_\text{par}} \bar{n_l}
.\end{equation}The $\langle \cdot \rangle$ operator corresponds here to the average amongst all the fragmentation outcomes that may occur within a structure of initial size $R_0$. Thus, for any scale $R$ there is on average a total amount of $\langle N_\text{tot}(R) \rangle$ children laying within the structure of size $R_0$ and $\langle N_\text{tot}(R) \rangle$ represents the whole collection of children of size $R$ produced by every single parent localised within their hosting structure. We define the fragmentation rate $\phi(R)$ as the variation of the fragments number per unit of logarithmic size:

\begin{equation}
    \phi(R) = - \dfrac{d \ln \langle N_\text{tot}(R) \rangle}{d \ln R}
    \label{eq:def_fragrate}
.\end{equation}The minus sign is set such that $\phi(R) > 0$ characterises an increase of the number of fragments at lower spatial scale (i.e. an actual fragmentation). With this definition, the fragmentation rate {statistic} represents the average variation of fragments number inside a multi-scale structure at each size $R$, rather than the local variation within a specific parent of size $R < R_0$. In addition, we can show that all the parents of scale $R_l \equiv R$ experiencing a fragmentation rate $\phi(R)$ produce on average $\langle \bar{n_l} \rangle$ fragments at the next scale $R_{l+1}$ according to

\begin{equation}
     \langle \bar{n_l} \rangle = \exp{\left[ - \int^{R_{l+1}}_{R_{l}} \phi(R') d\ln R' \right]}
     \label{eq:micro_n_connection}
.\end{equation}

The previous equation connects the {statistic} associated with the fragmentation rate $\phi(R)$ with the average multiplicity {statistic} of individual parents $\langle \bar{n_l} \rangle$. The fragmentation rate $\phi(R)$ is the first parameter of our model and aims to quantify the amount of fragments produced during the fragmentation cascade.

\subsubsection{Stellar density}

As fragmentation eventually ends at a stoping scale $R_\text{stop}$, we can compute the mean stellar density $\langle n_*(R_0) \rangle$ as the average number of newly born stars $\langle N_* \rangle_{R_0}$ contained within a region of any size $R_0$ per unit of volume,

\begin{equation}
    \langle n_*(R_0) \rangle = \dfrac{\langle N_* \rangle_{R_0}}{V(R_0)}
,\end{equation}

\noindent where $V(R_0)$ is the volume of the region. 

Assuming an object of scale $R_\text{stop}$ within a region of size $R_0$ produces a single star, then $\langle N_\text{tot}(R_\text{stop}) \rangle = \langle N_* \rangle_{R_0}$. Then, we can express the average number of newly born stars $\langle N_* \rangle_{R_0}$ as a function of the fragmentation rate $\phi(R)$ by integrating Eq. \ref{eq:def_fragrate},

\begin{equation}
    \langle N_* \rangle_{R_0} = N_\text{tot}(R_0) \exp{\left[ \int^{R_0}_{R_\text{stop}} \phi(R') dlnR' \right]}
    \label{eq:stellar_multiplicity}
,\end{equation}

\noindent where $N_\text{tot}(R_0) = 1$ because we compute the stellar multiplicity of a single structure of size $R_0$. 
Here the size $R_0$ of the region from which we average the number of young stars is to be considered as a variable while $R_\text{stop}$ is constant. The average number of newly born stars $\langle N_* \rangle_{R_0}$ is usefull to predict the multiplicity of stellar groups given their spatial extension (e.g. Sect.\ref{Sec:Number of YSOs in stellar groups}) The mean stellar density can also be used to compute the mean separation $\bar{d}$ between stars in a system as $\bar{d} \sim \langle n_* \rangle^{-1/3}$ regardless of their actual spatial distribution.


\subsection{Average mass of the fragments}
\label{sec:Average mass of the fragments}

\subsubsection{The mass transfer rate $\xi(R)$}

To obtain a second parameter based on the mass of the fragments, we introduce the fragment average formation mass efficiency $\langle \mathcal{E}(R) \rangle$ at any scale $R$ as

\begin{equation}
    \langle \mathcal{E}(R) \rangle = \dfrac{\langle M_\text{tot}(R) \rangle}{M_0}
,\end{equation}

\noindent {where} $M_0 = M(R_0)$ is the mass of the fragment at the initial scale $R_0$ and $\langle M_\text{tot}(R) \rangle$ is the average total mass of the fragments of scale $R$ resulting from every possible fragmentation outcome. Similarly as Eq.\ref{eq:def_fragrate} we define the mass transfer rate $\xi(R)$ as the log-derivative of $\langle \mathcal{E}(R) \rangle$ with respect to the size $R$:

\begin{equation}
    \xi(R) = ~ - ~~ \frac{d \ln \langle \mathcal{E}(R) \rangle}{d \ln R} = - \frac{d \ln \langle M_\text{tot}(R) \rangle}{d \ln R}
    \label{eq:def_masstransfer}
.\end{equation}

The mass transfer rate $\xi(R)$ quantifies the variation of the total mass that ends up in the whole collection of fragments of size $R$. If $\xi(R) < 0$ then the total mass contained in the fragments of size $R$ is lower than the mass of the fragments of size $R + dR$. {This means that the parents lose a fraction of their mass reservoir during the fragmentation process. The mass dispersal of a fragment into its environment can be done through outflows or radiative feedback \citep{kim2018} or any random turbulent motion that could unbind fluid particles from the fragment and partially destroy it (e.g. \citealt{heitsch2006, camacho2016, lu2020}). On the contrary}, if $\xi > 0$ the mass reservoir of the parents is smaller than the children reservoir. Thus, some mass is injected into the children, for example through accretion. Finally, if the mass of the parents is similar to the mass used to form the whole collection of children, we have $\xi = 0$. \\

The mass transfer rate $\xi(R)$ only takes into account the net outcome of the interplay between accretion and {mass dispersal}. Hence, the mass transfer rate does not discriminate the individual contributions of these two effects. Given the definition of the mass transfer rate $\xi(R)$ (Eq.\ref{eq:def_masstransfer}), we can express the average formation efficiency $\langle \mathcal{E}(R) \rangle$ as

\begin{equation}
    \langle \mathcal{E}(R) \rangle = \mathcal{E}_0 \exp{\left[ \int^{R_{0}}_R \xi(R') dlnR' \right]}
    \label{eq:efficaciteRxi}
,\end{equation}

\noindent with $\mathcal{E}_0 = 1$ as the total mass of fragments of size $R_0$ is equivalent to the mass $M_0$ of the initial structure. At any discrete scale $R_{l+1}$, we can write

\begin{equation}
    \langle \mathcal{E}(R_{l+1}) \rangle = \langle \mathcal{E}(R_{l}) \rangle \exp{\left[ \int^{R_{l}}_{R_{l+1}} \xi(R') dlnR' \right]}
.\end{equation}

Substituing $\langle \mathcal{E}(R_{l}) \rangle$ and $\langle \mathcal{E}(R_{l+1}) \rangle$ by their integral expression Eq.\ref{eq:efficaciteRxi}, we can express the average efficiency $\langle \epsilon_l \rangle$ between two discrete levels of fragmentation by

\begin{equation}
     \langle \epsilon_l \rangle = \exp{\left[ \int^{R_{l}}_{R_{l+1}} \xi(R') dlnR' \right]}
     \label{eq:micro_epsilon_connection}
.\end{equation}

The mass transfer rate $\xi(R)$ represents the second parameter of our model based on the total mass of the fragments produced in a given region. Its sign is determined by the dominant process between accretion and {mass-loss processes} (see Table \ref{tab:rate_ranges}).

Next, we consider the mass of a single fragment at scale $R$. Among all possible fragmentation outcomes, a randomly selected fragment has on average a mass $\langle M(R) \rangle$. This average mass is defined considering every object that might be produced within every fragmentation outcome. Hence, the expected mass $\langle M(R) \rangle$ one fragment have at scale $R$ considering every fragmentation outcome can be written as

\begin{equation}
    \langle M(R) \rangle = \frac{\langle M_{\text{tot}}(R) \rangle}{\langle N_\text{tot}(R) \rangle}
    \label{eq:average_mass}
,\end{equation}

\noindent that can be rewritten as

\begin{equation}
    \langle M(R) \rangle = M_0 \frac{\langle \mathcal{E}(R) \rangle}{\langle N_{\text{tot}}(R) \rangle}
    \label{eq:mass_noparam}
.\end{equation}

By taking the log-derivative of the Eq.\ref{eq:mass_noparam} and injecting the definition of both the fragmentation rate $\phi(R)$ (Eq.\ref{eq:def_fragrate}) and the mass transfer rate $\xi(R)$ (Eq.\ref{eq:def_masstransfer}) it can be shown that

\begin{equation}
    \frac{d \ln \langle M(R) \rangle}{d \ln R} = \phi(R) - \xi(R)
    \label{eq:mass_size_relation}
.\end{equation}

This relation implies that the mass efficiency of individual fragment takes into account two components. The first is given by the fragmentation rate $\phi(R)$ which is linked to the amount of children that have to share the same parental mass reservoir. The second is given by the mass transfer rate $\xi(R)$ that holds information concerning the amount of mass lost or obtained during the fragmentation process. 

\begin{table}
    \caption{Physical description of fragmentation rate $\phi$ and mass transfer rate $\xi$ as a function of their sign.}
    \label{tab:rate_ranges}
    \centering
    \begin{tabular}{l | c c c}
        \hline\hline
        & $> 0$ & $= 0$ & $< 0$ \\
        \hline
        \\
        \multirow{2}{*}{$\phi$} & Fragment & Monolithic &  Fragmentation \\
         & production & collapse & disruption \\
        \\
        \multirow{2}{*}{$\xi$} & Mass gain & Mass & Mass loss \\
        & (accretion) & conservation & (outflow, {dispersal}) \\
        \hline
    \end{tabular}
\end{table}

\subsubsection{Volumetric mass density}

Given the size $R$ of a fragment and its average mass $\langle M(R) \rangle$, one can compute the average mass density $\langle \rho(R) \rangle$ of spherical fragments as

\begin{equation}
    \langle \rho(R) \rangle = \dfrac{\langle M(R) \rangle}{k R^3}
,\end{equation}

\noindent where $k = 4\pi/3$ is the geometrical constant if the scale $R$ is defined as the radius of the structure while $k = \pi/6$ if it represents its diameter. By taking the log-derivation with respect to the size and using Eq.\ref{eq:mass_size_relation} we can show that

\begin{equation}
    \frac{d \ln \langle \rho \rangle}{d \ln R} = \phi(R) - \xi(R) - 3
    \label{eq:rho_size_relation}
.\end{equation}

The dependence of the average mass density with the fragmentation rate $\phi(R)$ and the mass transfer rate $\xi(R)$ implies that the fragmentation outcome is determined by the physical description of the fragments. In particular, a density fluctuation, induced for example by turbulence, can grow if the condition $\phi(R) - \xi(R) - 3 < 0$ is satisfied. In that case, this density enhancement generates a gravitational instability and the structure can locally collapses. If the internal heating and cooling processes of the collapsing structure is modelled with an equation of state (EOS), then the Eq.\ref{eq:rho_size_relation} may provide a mean to study the consequences of the EOS on fragmentation. We investigate in more details in Sect.\ref{Sect.Condition for a hierarchical fragmentation} the effect of a polytropic EOS on the fragmentation rate.


\subsection{Three modes of fragmentation}
\label{Sec:Three modes of fragmentation}

A structure may fragment hierarchically or collapse monolithically without fragmentation but it may also {disperse} under the influence of internal support. We evaluate the required conditions a parent needs to (i) fragment into several pieces, (ii) collapse to a single child or (iii) {disperse} into its environment. According to the definition of the fragmentation rate $\phi(R)$ (Eq.\ref{eq:def_fragrate}), a parent associated with $\phi = 0$ only forms on average a single child (see Sect.\ref{sec:section1-1-1}). The outcome of this situation is similar to that associated with a monolithic collapse (i.e. the formation of a single object)  \citep{shu_self_similar_1977}.
However, there may be other pairs $(\phi, \xi)$ that allow {single fragment formation}. \\

{To qualitatively investigate the conditions describing the different modes under which fragments can be formed,} we introduce the free-fall time $t_{\text{ff}}$ that represents the characteristic time with which a gas structure collapses under its own gravity. {We consider that a} child emerging in a parent whose free-fall time is $t_{\text{ff},l}$ {exists} if its own free-fall time $t_{\text{ff},l+1}$ is smaller than that of its parent. {If the parent collapses faster than its children, the net product of fragmentation is equivalent to a monolithic collapse with virtual children that could merge during the collapse of the parent.} Hence, a child exists if

\begin{equation}
    \frac{t_{\text{ff},l+1}}{t_{\text{ff},l}} < 1
.\end{equation}

For spherical fragments the free-fall time depends on the mass density $\rho$ such that

\begin{equation}
    t_{\text{ff}} = \sqrt{\dfrac{3 \pi}{32 G \rho}} \propto \rho^{-1/2}
.\end{equation}

Then, a child exists within its parent as an actual fragment if the parent density is lower than the density of the possible child:

\begin{equation}
    \frac{\rho_{l}}{\rho_{l+1}} < 1
.\end{equation}



As a consequence, the density of the fragments needs to increase as fragmentation proceeds, which leads to the following condition for the existence of the children according to Eq.\ref{eq:rho_size_relation}:

\begin{equation}
    \xi(R) + 3 > \phi(R)
.\end{equation}

\noindent By definition hierarchical fragmentation also needs $\phi(R) > 0$ such that multiple fragments are formed (Eq.\ref{eq:def_fragrate}) and we get the condition for hierarchical fragmentation {with children of increasing density}:

\begin{equation}
    \xi + 3 > \phi > 0
    \label{eq:hierarchical_condition}
.\end{equation}


Then, we {investigate the hierarchical fragmentation process ($\phi > 0$) under which children are formed with decreasing density (i.e. when their parent free-fall time is higher)}. First, we consider

\begin{equation}
    \phi \geqslant \xi + 3 \geqslant 0
    \label{eq:not_hierarchical}
.\end{equation}

\noindent {In this case, the children collapse less rapidly than their parent as they are more diffuse. The children material disperse and the parental structure collapses, so the net product of fragmentation is the formation of a single object. The children formed through fragmentation are called virtual as they do not exist anymore after one parental free-fall time. Since the children are virtual, the outcome of this condition can be modelled using an effective fragmentation rate $\phi_\text{eff} = 0$. This scenario implies that the parental structure is not necessarily substructured across the scales for all $\phi > 0$ but may be observed as collapsing monolithically with dispersed children (red structure in Fig. \ref{fig:modes_fragmentation}).}


Finally, we consider the condition

\begin{equation}
    \phi \geqslant 0 > \xi + 3
    \label{eq:not_hierarchical}
,\end{equation}

\noindent under which hierarchical fragmentation is also prevented and the mass transfer rates satisfies $\xi < -3$. {In this condition, the children also collapse less rapidly than their parent as they are more diffuse. Nonetheless, the mass loss by the collapsing parent is not compensated by its volume reduction so its density also decreases as it contracts (Eq.\ref{eq:rho_size_relation} with $\phi_\text{eff} = 0$). Therefore, any structure with $\xi < -3$ eventually disperses into its environment with its virtual children as their density decreases. The net product of fragmentation at smaller scale is then equivalent to a fragmentation disruption ($\phi < 0$) as fragments progressively disappear throughout the spatial scales (black structure in Fig. \ref{fig:modes_fragmentation}).}

\begin{figure}
    \centering
    \includegraphics[width=8cm]{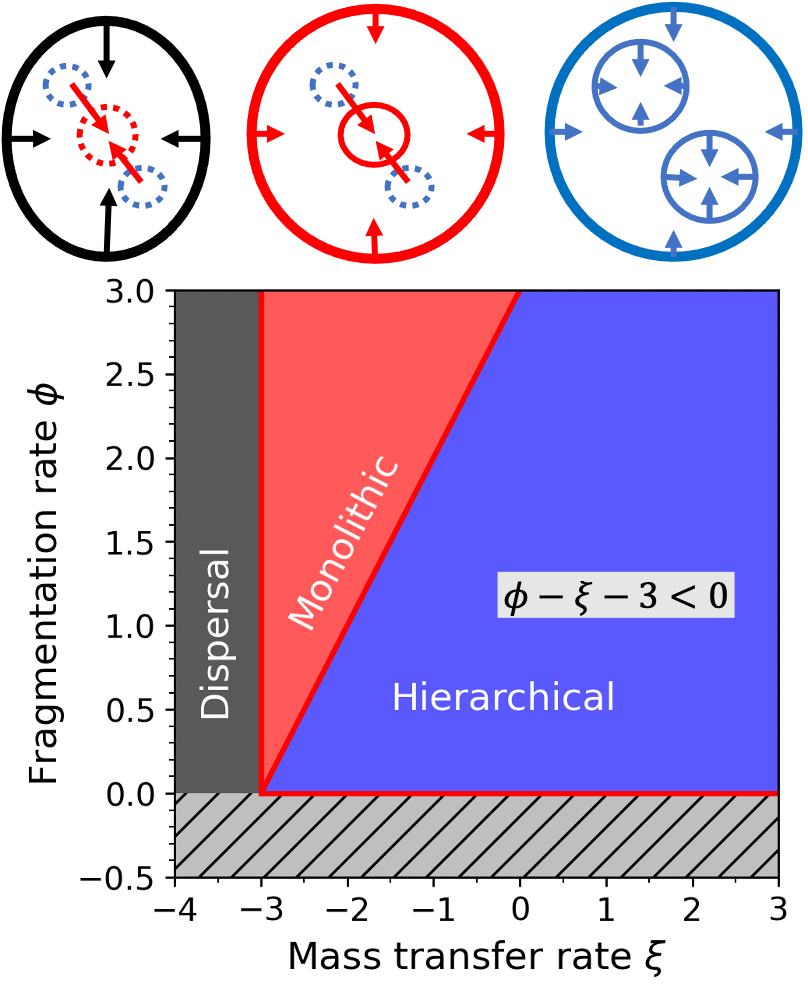}
    \caption{Different fragmentation modes in a fragmentation rate $\phi$ {vs} mass injection rate $\xi$ diagram.
    The hierarchical, monolithic, and {mass dispersal} modes are shown respectively as blue, red, and black {areas and each respective sketch is above the figure. When $\phi - \xi - 3 < 0$, the children collapse before their parent (hierarchical). At lower $\xi$ the children do not have time to collapse before their parent does and a single fragment is effectively formed (monolithic). For $\xi < -3$ this single fragment progressively disperses as it gets more and more diffuse during the collapse (dispersal). In the hatched region $\phi < 0$, fragmentation is disrupted}.}
    \label{fig:modes_fragmentation}
\end{figure}

In a nutshell, we distinguish three fragmentation modes (Fig. \ref{fig:modes_fragmentation}) in the case of $\phi > 0$ with this qualitative study based on the free-fall time: 

\begin{itemize}
    \item {a dispersal mode if $\xi < -3$ as the children and the parent densities decrease with their respective collapse};
    \item {an effective monolithic collapse if $\phi \geqslant \xi + 3 \geqslant 0$ as the children disperse into their denser collapsing parent};
    \item {a hierarchical fragmentation mode if $\xi(R) + 3 > \phi(R) > 0$ as the children are denser than their parental structure}.
\end{itemize}

In addition, we also recall that {fragmentation is disrupted} for $\phi < 0$ while the monolithic {collapse} can be represented by $\phi = 0$.

It is important to note that the modes of fragmentation we extracted are based on the qualitative study of the free-fall time that does not account for thermal nor turbulent support. As turbulent support is more effective for larger structures (see Sect.\ref{Sec:Turbulent energy cascade}), their free-fall time should be higher than the free-fall time of their children. In that case, the hierarchical condition should be less restrictive {(}i.e. some of the $\phi \geqslant \xi + 3 \geqslant 0$ solutions could be hierarchical{)}. Similarily for thermally supported fragments, if the children have a higher internal pressure than their parents (for example because of opacity heating), we may expect the hierarchical condition to be more restrictive and some of the $0 < \phi < \xi + 3$ solutions may not be hierarchical.


\section{Hierarchical model applied to a gravo-turbulent framework}
\label{sec:section2}

We introduced in Sect.\ref{sec:section1} the general structure of our fragmentation model. Hereafter we implement our model into a physical background using a gravo-turbulent framework (Sect.\ref{Sec:Theoretical framework}) in order to find analytical solution of the fragmentation rate $\phi(R)$ and determine the conditions that allow a cloud to hierarchically fragment (Sect.\ref{Sec:An analytical solution for the fragmentation rate}). Then, we investigate the implications of this gravo-turbulent fragmentation on how fragmentation behave with respect to the spatial scales (Sect.\ref{sec:Variations of the fragmentation rate}).


\subsection{Theoretical framework}
\label{Sec:Theoretical framework}

\subsubsection{Turbulent energy cascade}
\label{Sec:Turbulent energy cascade}

Turbulent kinetic energy cascades along spatial scales \citep{kolmo1941}. The power spectrum of turbulent energy is scale-free such that the resulting fragmentation process is inherently hierarchical until the turbulent cascade is terminated. In that perspective, we may assume that the turbulent velocity of molecular clouds $V_\text{rms}$ could be modelled with any function of the spatial scale so that we can introduce an exponent $\eta(R)$ characterising the local variations of $V_\text{rms}(R)$ {as}

\begin{equation}
    \eta(R) = \dfrac{d \ln V_\text{rms}}{d \ln R} 
.\end{equation}

If such a function exists, the turbulent velocity $V_\text{rms}(R)$ is scale-free if $\eta(R)$ is constant, which implies

\begin{equation}
    V_\text{rms} = V_0 \left( \frac{R}{R_0} \right)^\eta
    \label{eq:vitesserms}
.\end{equation}

\noindent This relation is similar to the  Larson law \citep{larson1981_relation} for non-thermal motions with $V_0 = 1~\text{km.s}^{-1}$, $R_0 = 1$~pc and $\eta = 0.38$, while $\eta = 1/3$ corresponds to a Kolmogorov cascade \citep{kolmo1941}. Here the scale $R$ corresponds to the length of the cloud. This scale-free relation has been interpreted as the consequence of a cloud remaining in virial equilibrium but may also arise from a Kolmogorov cascade of turbulent energy \citep{kolmo1941}. Although Larson's law is empirically valid for sizes $R > 0.1$~pc \citep{henne&falga}, we assume in the following that Eq.\ref{eq:vitesserms} remains valid for sizes $R < 0.1$~pc. This assumption is reasonable at large scales but may be questionable at smaller scales. This assumption neglects the additional non-thermal support on the fragmentation outcome at smaller scale, {for example the} magnetic support. In fact, the Eq.\ref{eq:vitesserms} may not be verified for non-virialised collapsing structures \citep{traficante2018}. As pressure support could be underestimated, we expect to overestimate the number of fragments produced with our model (e.g. Sect.\ref{section:additional support}). As long as the function $V_\text{rms}(R)$ exists, it could be modelled with any $\eta(R)$ and could be adapted for higher non-thermal support by changing the value of the power index $\eta$. The choice of using the scale-free Larson law should be considered as purely illustrative as we could assess the influence of higher velocities on the fragmentation properties.

\subsubsection{$\rho$-PDF induced by turbulence}
\label{Sec:rhoPDF}

In turbulent flow, the density $\rho$ of a fluid particle contained in a cloud of mean density $\bar{\rho}$ fluctuates stochastically. In a gravo-turbulent fragmentation, the random field of density fluctuations in an isothermal medium has a log-normal {statistic} \citep{passot1998, padoan2002}. Thus, the density contrast $\delta = \ln (\rho / \bar{\rho})$ follows a normal distribution $P(\delta)$:

\begin{equation}
 P(\delta) = \frac{1}{\sqrt{2\pi}\sigma} \exp\left[{- \frac{\left(\delta + \frac{\sigma^2}{2}\right)^2}{2\sigma^2}}\right]
 \label{eq:logrho_PDF}
,\end{equation}

\noindent with a variance $\sigma^2 = \ln(1 + b^2\mathcal{M}^2)$ where $b^2 \approx 0.25$ and $\mathcal{M}$ is the turbulent Mach number of the cloud.

The choice of describing the field of density fluctuations with a normal distribution is a valid solution for an isothermal supersonic medium. Under the assumption that turbulent velocity can be modelled by the Larson law (see Sect.\ref{Sec:Turbulent energy cascade}), the former condition is always satisfied for spatial scales above $R \sim 0.1$~pc at cloud temperature $T \sim 10$~K. Below $R \sim 0.1$~pc, adiabatic conditions arise when the fragments are dense enough such that they become opaque to their own radiation. At these small scales the turbulence is subsonic according to the Larson law. In such a subsonic regime the amplitude of density fluctuations is reduced due to the low level of turbulence. According to \cite{passot1998}, we can consistently employ the log-normal $\rho$-PDF given Eq.\ref{eq:logrho_PDF} under this small density fluctuations approximation. 

A collapsing structure develops over time a powerlaw radial density profile $\rho(r) \propto r^{-2}$ where $r$ is the distance to the centre of the structure \citep{shu_self_similar_1977}. This radial profile may bias the $\rho$-PDF since the densest regions concentrate within the centre of a structure while the edges are more diffuse. As a consequence, the $\rho$-PDF develops a power law tail at high densities \citep{kritsuk2011_rhopdf}. Therefore, this $\rho$-PDF would no longer be log-normal and one should consider the density profile of a collapsing structure. However, our aim is to evaluate the condition for an instability to develop in the future. If a structure is characterised by this radial profile, the central instability has already grown and no additional fluctuation has appeared in the meantime to disrupt the emergence of a $\rho(r) = r^{-2}$ profile. If other fluctuations have appeared in the meantime, subfragments may actually exist. For a more detailed description of this mechanism, we would have to consider the dynamics of the formation of fragments which goes beyond the scope of our quasi-static model.

\subsubsection{Collapse threshold}

A random density fluctuation may lead to a local collapse within the cloud if the resulting density is sufficient to trigger a gravitational instability. The critical value $\delta_c$ beyond which such an instability can grow has been formally expressed by \cite{H&C2008} in the framework of Jeans' instability considering the turbulent support of a non-magnetised isothermal star-forming collapsing structure

\begin{equation}
    \delta_c = \ln \left[\frac{\pi^{5/3}}{6^{2/3}} ~ \frac{3 C_s^2 + V_\text{rms}^2}{3G R^2 \bar{\rho}} \right]
    \label{eq:delta_c}
,\end{equation}

\noindent where $G = 6.67 \times 10^{-11}~\text{m}^3\text{kg}^{-1}\text{s}^{-2}$ is the gravitational constant and $C_s = (k_B T/\mu m_H)^{1/2}$ is the sound velocity in which $k_B \approx 1.38 \times 10^{-23}$~J.K$^{-1}$ the Boltzmann constant, $T$ is the cloud mean temperature, $\mu = 2.33$ the mean molecular weight of the cloud and $m_H = 1.67\times 10^{-27}$~kg the hydrogen mass. Then, a volume $V(R)$ contained in a larger structure collapses if a fluctuation $\delta \geqslant \delta_c$ arises. The probability $\mathcal{P}(\delta \geqslant \delta_c)$ for this volume to collapse is then characterised by the complementary cumulative distribution of the PDF $P(\delta)$. Considering the $\rho$-PDF given by Eq.\ref{eq:logrho_PDF}, we obtain after integration

\begin{equation}
    \mathcal{P}(\delta \geqslant \delta_c) = \frac{1}{2} \left( 1 - \text{erf} \left[ \frac{\delta_c + \frac{\sigma^2}{2}}{\sigma \sqrt{2}} \right] \right)
    \label{eq:cumulative_rho_pdf}
,\end{equation}

\noindent where erf(x) is the error function. 

In the following, fragmentation is considered to be the result of the local conditions within the cloud. Thus, we assume that the variance $\sigma(R)^2 = \ln(1 + 0.25\mathcal{M}(R)^2)$ is a function of the spatial scale $R$ through the local mach number of the fragment we consider. This local Mach number is defined as the ratio between the turbulent velocity $V_\text{rms}(R)$ and the sound velocity $C_s$. Hence, the complementary cumulative function of Eq.\ref{eq:cumulative_rho_pdf} represents the probability that any unstable fluctuation of size $R$ emerges within a structure of density $\bar{\rho}$ with a variance $\sigma^2$ determined by its local mean Mach number.

\subsubsection{Equation of state}

During the collapse of a star-forming structure, the fragments density eventually reaches an opacity limit that induces internal heating and increases the fragment thermal pressure. This heating can be modelled with an EOS regulated by an adiabatic index $\gamma$ that describes the heating rate of a fragment \citep{masunaga1998}. Here the index $\gamma$ represents the heat capacities ratio of a specific chemical specie. Adiabaticity is reached when the fragment density $\rho$ reaches the density $\rho_L \sim 10^{-13}$~g/cm$^{-3}$ of the first Larson core \citep{larson1969_fcore}. For molecular hydrogen, the adiabatic index $\gamma = 5/3$ for temperatures $T < 100$~K. The relationship between the temperature $T$ of a structure and its density $\rho$ is described by the following EOS \citep{Lee&H2018_adiafirstlarson}:

\begin{equation}
T(\rho) = T_0 \left[ 1 + \left(\frac{\rho}{\rho_L}\right)^{\gamma - 1} \right]
    \label{eq:adia_EOS}
,\end{equation}

\noindent with $T_0 \approx 10$~K for a dense clump. We also define the local polytropic EOS at scale $R$

\begin{equation}
    \dfrac{d \ln T}{d \ln \rho} = p - 1
    \label{eq:local_EOS}
,\end{equation}

\noindent where $p$ is the polytropic index describing locally the heating rate of a fragment of density $\rho(R)$ at scale $R$. Thus, for densities $\rho > \rho_L$, we have asymptotically $T \propto \rho^{\gamma-1}$ with a local variation modelled with $p = \gamma$. On the contrary, for densities $\rho < \rho_L$ temperature is constant and we have $p = 1$ {(}i.e. an isothermal EOS{)}. The sound speed $C_s$ is a function of temperature $C_s(T)$ such that $C_s$ depends on the density of the fragments. Thus, the collapsing criteria $\delta_c$ can be expressed as a function of size $R$ and density $\rho$ only.


\subsection{An analytical solution for the fragmentation rate $\phi(R)$}
\label{Sec:An analytical solution for the fragmentation rate}

In the following, we assume an isotropic fragmentation, meaning that the probability of finding fragments in specific parts of a parent does not depend on their location within their parent. Since magnetic field usually breaks spherical symmetries, we consider non-magnetised collapsing structures. We also consider the fragmentation as an ergodic process such that the evolution over time of both $\phi(R)$ and $\xi(R)$ yields to the same variations as the average variation taken over a whole cloud. We {also} consider that fragmentation only depends on the local properties of the parents such that they do not interact with each other. {Finally, we assume that no fragments can be formed in the inter-fragments media at scale $R_{l+1}$ without a direct parental structure at scale $R_l$ so fragmentation is entirely hierarchical.}

\subsubsection{Procedure to compute the fragmentation rate}
\label{Sect:Procedure to compute the fragmentation rate}

Our aim is to compute the fragmentation rate $\phi(R)$ using the $\rho$-PDF we introduced in Sect.\ref{Sec:rhoPDF}. The maximum number of fragments a structure can contain corresponds roughly to its filling factor $V_0/V(R)$, where $V(R)$ is the volume of a child with characteristic size $R$. 
Considering a fluctuation $\delta(R)$, each one of these children has a probability $\mathcal{P}(\delta \geqslant \delta_c)$ of collapsing. The average number of fragments of size $R$ is then

\begin{equation}
    \langle N(R) \rangle = N(R_0) \left( \frac{R}{R_0}\right)^{-3} \mathcal{P}(\delta \geqslant \delta_c)
    \label{eq:generalnumber}
,\end{equation}

\noindent where $N(R_0) = 1$ as we start from a single cloud. To obtain the fragmentation rate $\phi(R)$, we use its definition given by Eq.\ref{eq:def_fragrate} and take the log-derivative of the equation above with respect to the logarithm of the size{, so we obtain}

\begin{equation}
    \phi(R) = 3 - \frac{d \ln \mathcal{P}(\delta \geqslant \delta_c)}{d \ln R}
    \label{eq:generalphi}
.\end{equation}

This expression can be computed analytically for any $\rho$-PDF provided it is differentiable. Since $\delta_c$ depends on the average density $\bar{\rho}$ (Eq.\ref{eq:delta_c}) of the parents, it is necessary to solve simultaneously Eq.\ref{eq:rho_size_relation} which describes the variation of the average density of the fragments. Provided we set the initial conditions on the size $R_0$ of a structure and its initial mean density $\bar{\rho_0}$ to define a starting collapsing threshold $\delta_c(R_0, \rho_0)$, we can compute the fragmentation rate by solving the following coupled system:

\begin{equation}
    \begin{cases}
        \phi(R, \bar{\rho}) = 3 - \dfrac{d \ln \mathcal{P}\big(\delta \geqslant \delta_c(R, \bar{\rho})\big)}{d \ln R} \\[15pt]
        \dfrac{d \ln \bar{\rho}}{d \ln R} = \phi(R, \bar{\rho}) - \xi(R) - 3
    \end{cases}
.\end{equation}

The analytical derivation of the fragmentation rate $\phi(R, \bar{\rho})$ under all of our considerations (see Sect.\ref{Sec:Theoretical framework}) is given in Appendix \ref{Appendix:computefrag}. Expanding Eq.\ref{eq:generalphi} we find that

\begin{equation}
        \phi(R, \bar{\rho}) = \dfrac{3}{1 - A_\text{th}} - \bigg( \xi(R) + 3 \bigg) \dfrac{A_\text{th}}{1 - A_\text{th}} + \dfrac{2A_\text{turb}}{1 - A_\text{th}}
\label{eq:analytic_fragmentation_rate}
,\end{equation}

\noindent where $A_\text{th}(R, \bar{\rho}), A_\text{turb}(R, \bar{\rho})$ are associated with the thermodynamics of the collapse (i.e. the EOS) and the turbulent energy cascade (i.e. in this case the scale-free Larson law) respectively {, with}

\begin{equation}
    \begin{cases}
        A_\text{th} = A_\text{p} \left[ 1 - \dfrac{d \ln T}{d \ln \rho} \bigg( \hat{C_s}^2 + \dfrac{v \hat{\mathcal{M}}^2}{\sqrt{2} \sigma} \bigg) \right] \\[15pt]
        A_\text{turb} = A_\text{p} \left[1 -  \eta \bigg( \hat{V}_\text{rms}^2 - \dfrac{v \hat{\mathcal{M}}^2}{\sqrt{2} \sigma}\bigg) \right] \\[15pt]
        A_\text{p} = - \dfrac{\exp[-u^2]}{\mathcal{P}\big(\delta \geqslant \delta_c\big)\sigma \sqrt{2\pi}} 
    \end{cases}
,\end{equation}

\noindent where $A_\text{p}(R, \bar{\rho})$ ponderates both $A_\text{th}(R, \bar{\rho})$ and $A_\text{turb}(R, \bar{\rho})$ with the fluctuation {statistic}. The reduced variables $\hat{C_s}^2, \hat{V}_\text{rms}^2, \hat{\mathcal{M}}^2$ are given by

\begin{equation}
    \begin{cases}
    \hat{C_s}^2 = \dfrac{3C_s^2}{3C_s^2 + V_\text{rms}^2} \\[15pt]
    \hat{V}_\text{rms}^2 = \dfrac{V_\text{rms}^2}{3C_s^2 + V_\text{rms}^2} \\[15pt]
    \hat{\mathcal{M}}^2 = \dfrac{b^2 \mathcal{M}^2}{1 + b^2 \mathcal{M}^2}
    \end{cases}
,\end{equation}

\noindent while the parameters $u$ and $v$ are reduced centred variables:

\begin{equation}
    \begin{cases}
        u(R, \bar{\rho}) = \dfrac{\delta_c(R, \bar{\rho}) + \sigma(R)^2/2}{\sigma(R) \sqrt{2}} \\[15pt]
        v(R, \bar{\rho}) = \dfrac{\delta_c(R, \bar{\rho}) - \sigma(R)^2/2}{\sigma(R) \sqrt{2}}
    \end{cases}
.\end{equation}

The fragmentation rate expressed in Eq.\ref{eq:analytic_fragmentation_rate} consists of the addition of three terms. From left to right, the first one reflects both the isotropy of fragmentation and the spheroidal geometry of the fragments we consider. The second one characterises the dependancy of $\phi(R)$ with the mass transfer rate $\xi(R)$ since the density of fragments inherently couples them. The third term is associated with the turbulent energy cascade that contributes only in the supersonic limit as this term cancels out in a subsonic flow. All these terms are weighted by both the density fluctuation {statistic} and the local EOS. In particular, as the fragmentation rate sensitively depends on the EOS, heating processes are a dominant and determining factors in the fragmentation outcomes as we show in Sect.\ref{Sec:Natural stop for the fragmentation}. 

In Eq.\ref{eq:analytic_fragmentation_rate} the mass transfer rate $\xi(R)$ can be represented by any function. Without any suitable model to predict $\xi(R)$, we assume that $\xi$ is a constant so that we can test in the following different fragmentation outcomes as a function of the mass transfer rates. We have for the mean efficiency

\begin{equation}
    \langle \mathcal{E}(R) \rangle = \mathcal{E}_0 \left( \dfrac{R}{R_0} \right)^{-\xi}
    \label{eq:scale_free_eff}
,\end{equation}

\noindent with $\mathcal{E}_0 = 1$ as the total mass of fragments of size $R_0$ is equivalent to the mass $M_0$ of the initial structure. 

\subsubsection{Condition for a hierarchical fragmentation}
\label{Sect.Condition for a hierarchical fragmentation}

In Sect.\ref{Sec:Three modes of fragmentation} we established qualitatively different fragmentation modes depending on the numerical values of both the fragmentation rate $\phi$ and mass transfer rate $\xi$ (see Fig. \ref{fig:modes_fragmentation}). We can compute analytically using Eq.\ref{eq:analytic_fragmentation_rate} the solutions that determine these modes under subsonic and supersonic conditions. A parent hierarchically fragments if the condition $\xi + 3 > \phi(R) > 0$ is satisfied. In the subsonic limit in which $C_s \gg V_\text{rms}$, we can show (see Appendix \ref{Appendix:asymptotic}) that

\begin{equation}
    \phi(R, \bar{\rho}) \approx \xi + 3 - \frac{2 \bigg( 1 + \eta \dfrac{\delta_c}{2} \bigg)}{ 1 - \dfrac{d \ln T}{d \ln \rho} \bigg( 1 + \dfrac{\delta_c}{2} \bigg)}
    \label{eq:subsonic_fragrate}
.\end{equation}

To simplifiy Eq.\ref{eq:subsonic_fragrate} we assume that the mass reservoir is conserved between the scales so that $\xi = 0$. We also assume marginally unstable parents such that $\rho_c \sim \bar{\rho}$ and $\delta_c \sim 0$. Then, using the local EOS (see Eq.\ref{eq:local_EOS}), Eq.\ref{eq:subsonic_fragrate} can be simplified to

\begin{equation}
    \phi \approx \dfrac{4 - 3p}{2 - p}
    \label{eq:phi_adiabatic_subsonic}
.\end{equation}

\noindent In these conditions, hierarchical fragmentation occurs when the condition $3 > \phi(R) > 0$ is satisfied, {which means} when the polytropic index $p$ satisfies

\begin{equation}
        p < \frac{4}{3}
.\end{equation}

In order to collapse monolithically we can also show that the condition $\phi = 0$ implies (under the same approximations) that $p = 4/3$. Beyond $p = 4/3$, the children produced {disperse their mass}. This result is similar to the traditional analysis of Jeans fragmentation that derives a critical Jeans mass $M_J \propto \rho^{\frac{3}{2} (p-\frac{4}{3})}$. Hence, $p = 4/3$ is a critical point below which the Jeans instabilities grow as the fragments contract quasi-statically. The Eq.\ref{eq:subsonic_fragrate} generalises the Jeans analysis in the subsonic limit by taking into account the amount of mass added or removed from the fragments ($\xi$), the stability of the collapsing structure ($\delta_c$) but also the decay of the turbulent cascade ($\eta$). 

In the supersonic limit ($C_s \ll V_\text{rms}$), $\phi \xrightarrow[]{} 3$ for any $\xi$. Hence, any loss of mass $\xi < 0$ during the fragmentation may likely induce a monolithic collapse as the pair $(\phi, \xi)$ would satisfy the monolithic condition $\phi \geqslant \xi + 3 \geqslant 0$ (see Fig. \ref{fig:modes_fragmentation}). 


\subsection{Variations of the fragmentation rate}
\label{sec:Variations of the fragmentation rate}

\subsubsection{Turbulent vs thermal support}
\label{sec:Turbulent vs thermal support}

\begin{figure}
    \centering
    \includegraphics[width=8cm]{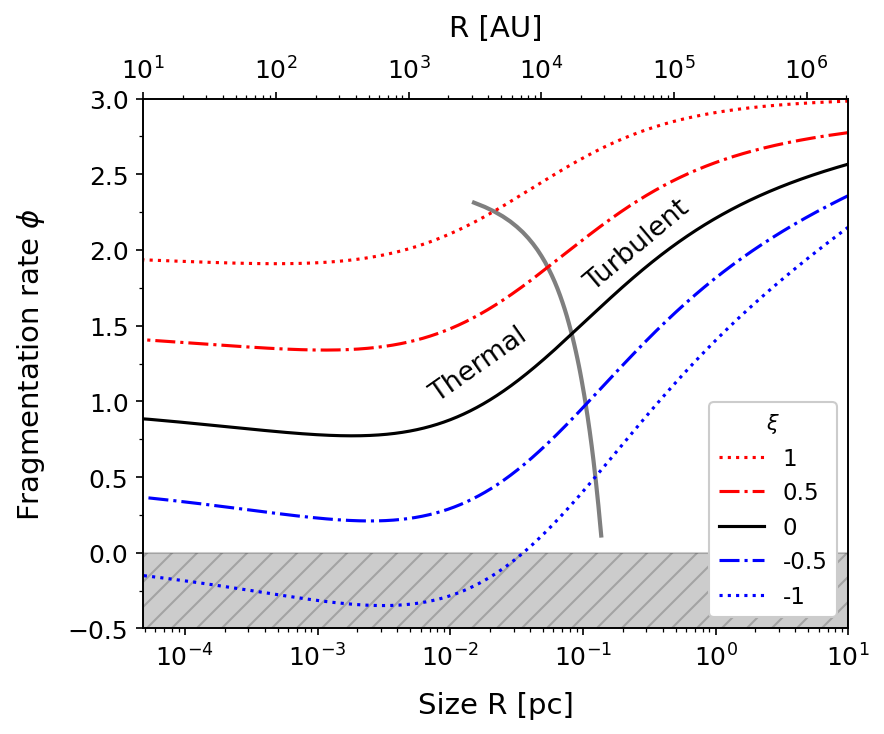}
    \caption{Fragmentation rate $\phi$ as a function of spatial scale $R$ for isothermal EOS and different values of mass transfer rate $\xi$ with cloud initial size $R_0 = 10$~pc and initial density $\rho_0 = 10^{-21}$~g/cm$^{3}$. The hatched area represents the $\phi \leqslant 0$ space in which there is no hierarchical fragmentation. The grey vertical line separates turbulent (supersonic) fragmentation from thermal (subsonic) fragmentation.}
    \label{fig:isothermal_phi}
\end{figure}

\begin{figure}
    \centering
    \includegraphics[width=8cm]{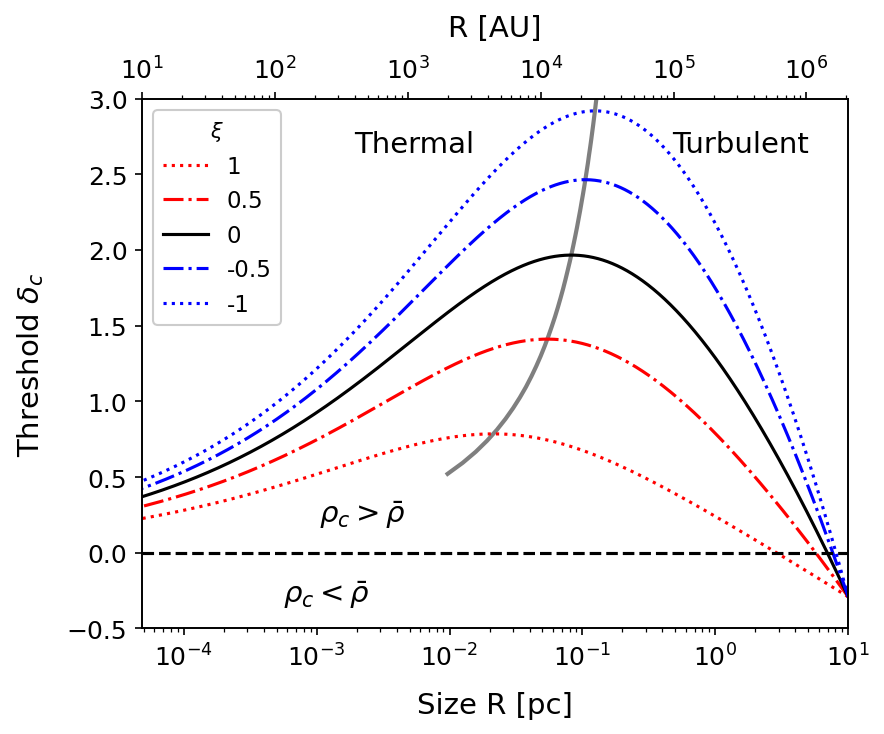}
    \caption{Contrast density threshold $\delta_c = \ln(\rho_c/\bar{\rho})$ as a function of spatial scale $R$ for isothermal EOS and different values of mass transfer rate $\xi$ with cloud initial size $R_0 = 10$~pc and initial density $\rho_0 = 10^{-21}$~g/cm$^{3}$. The grey vertical line separates turbulent (supersonic) fragmentation from thermal (subsonic) fragmentation.}
    \label{fig:threshold_profile}
\end{figure}

To assess the variations of the fragmentation rate $\phi(R)$ with the size $R$ of the fragments we consider in the following a cloud of size $R_0 = 10$~pc and mass $M_0 = 10^4~\text{M}_\odot$, {so the} initial mean density $\bar{\rho_0} \sim 10^{-21}$~g/cm$^3$ with a cloud radius $R_0/2$. The fragmentation rate is computed using an isothermal EOS with $T = 10$~K and for different mass transfer rates $\xi = \text{constants}$ varying between 1 and -1 (Fig. \ref{fig:isothermal_phi}). 

We distinguish two regimes of fragmentation. The first one is a fragmentation driven by turbulence at large scales while the second one is driven by gravity at smaller scales. The transition between these two regimes occurs for spatial scale around $R \sim 0.1$~pc that is generally designated as the characteristic sonic length \citep{vazquez2003_sonicscale, ballesteros2004, Palau2015} from which turbulence becomes negligible compared to thermal support.

In the framework of our model, this transition can be understood as the moment when the collapse threshold $\delta_c$ reaches a maximum (Fig. \ref{fig:threshold_profile}). Since $\delta_c < 0$ at $R_0$, the cloud is initially unstable (see Fig. \ref{fig:threshold_profile}). The threshold $\delta_c$ eventually becomes positive while the turbulence is still supersonic. If $R > 0.1$~pc then $\delta_c$ increases for decreasing $R$ and it is more difficult for a fluctuation to be gravitationnally unstable because of strong turbulence support. Once $R < 0.1$~pc, turbulent support has sufficiently decayed not to be dominant anymore compared to thermal support. Hence, the threshold $\delta_c$ decreases as $R$ decreases since a random fluctuation is more and more likely to induce a local collapse. The maximum of the function $\delta_c(R)$ for which this transition occurs can be analytically computed by taking the derivative with respect to the variable $\ln(R)$. Then we can show that this derivative cancels out when

\begin{equation}
   \dfrac{2 \eta \mathcal{M}^2}{3 + \mathcal{M}^2} = \phi - \xi - 1
   \label{eq:condition_sonic}
.\end{equation}

For the values of $\xi$ we considered, this transition occurs when $\phi - \xi \approx 1.1 - 1.4$. Using the analytical solution given Eq.\ref{eq:condition_sonic} for $\eta = 0.38$, we find that super-to-subsonic transition occurs when $\mathcal{M} \approx 0.7 - 1.8$ such that the Mach number needs to be of the order of unity.

In the supersonic limit, the fragmentation rate asymptotically tends to $\phi \approx 3$ (see Sect.\ref{Sect.Condition for a hierarchical fragmentation}) while in the subsonic limit the fragmentation rate is defined by Eq.\ref{eq:subsonic_fragrate}. In addition, the fragmentation is scale-free if $\phi$ is constant, which is the case in the supersonic limit, as we may expect from a scale-free turbulence cascade. This transition between supersonic and subsonic fragmentation necessarily breaks the turbulent scale-free fragmentation. In the subsonic case, the fragmentation is scale-free if the structure is marginally unstable (Eq. \ref{eq:phi_adiabatic_subsonic}) and if $\xi$ is constant. In fact, the fragmentation rate varies slowly in the subsonic regime for $R < 0.01$~pc (Fig. \ref{fig:isothermal_phi}) and can be considered as locally scale-free.

Nonetheless, the scale-free subsonic fragmentation cannot be reached for all mass transfer rate. If $\xi < -0.5$ fragmentation falls {into a disrupted} regime as $\phi(R) < 0$ (Fig. \ref{fig:isothermal_phi}). In that case, fragmentation stops at scale $R_\text{stop} > 10^{-2}$~pc. Using Eq.\ref{eq:scale_free_eff} we find that fragmentation is prevented if the last fragments are produced with a mass efficiency $\langle \mathcal{E}(R_\text{stop}) \rangle < 3\%$. Hence, we should expect large-scale mass efficiencies to be higher than the $3\%$ order of magnitude in order to produce star-forming structures.

\subsubsection{Self-regulation of the fragmentation}
\label{Sec:Self-regulation of the fragmentation}

\begin{figure*}[!t]
    \centering
    \includegraphics[width=18cm]{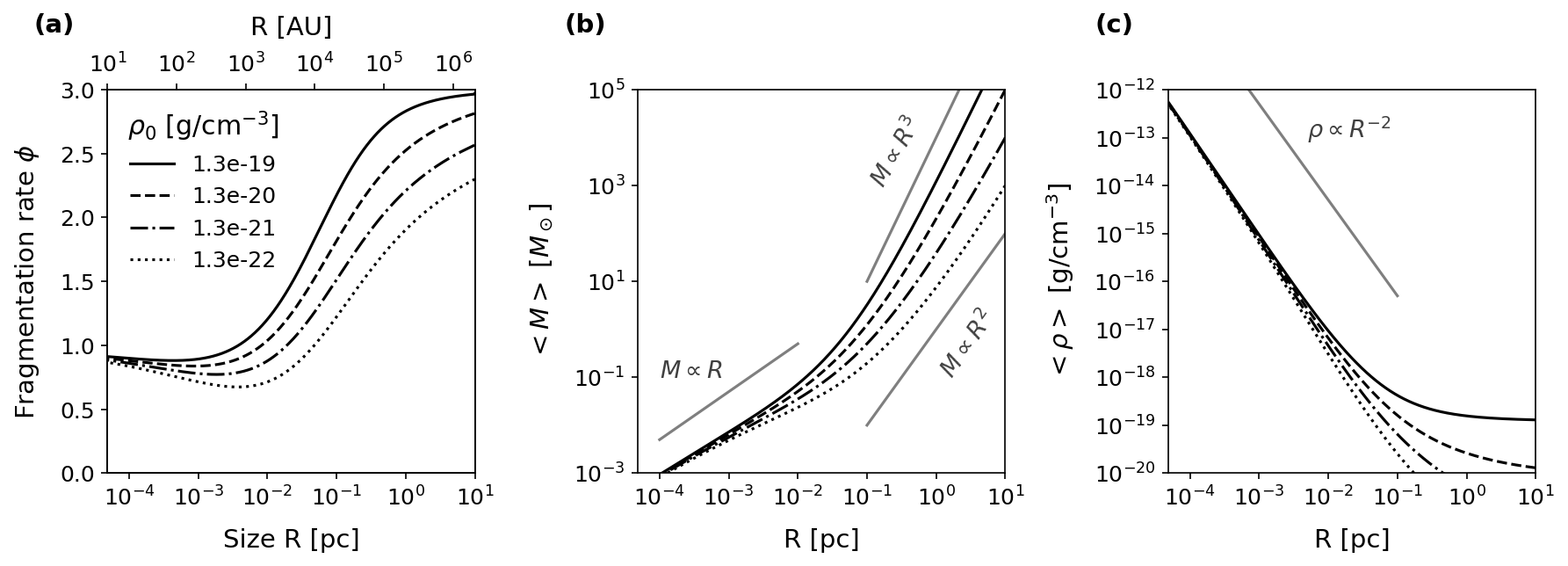}
    \caption{Variations of the fragmentation rate $\phi$ (a), {mean fragment mass $\langle M \rangle \propto R^{\phi - \xi}$ (b), and mean fragment density $\langle \rho \rangle \propto R^{\phi - \xi - 3}$} (c) as a function of their spatial scale $R$ for different initial cloud densities $\rho_0$ with an isothermal EOS {using $\xi = 0$}. The legend of (a) is common to the three plots. The scaling relation $M \propto R$ in (b) indicates the asymptotic limits of a marginally unstable Bonnor--Ebert structure \citep{Bonnor1956}. The scaling relation $\rho \propto R^{-2}$ in (c) represents the density profile of a self-gravitating isothermal sphere \citep{shu_self_similar_1977}.}
    \label{fig:initial_condition}
\end{figure*}

We may check the influence of the cloud's initial conditions on the evolution of the fragmentation rate $\phi(R)$. Since the important physical property of a structure for the fragmentation is its mean density $\bar{\rho}$, we choose to vary the initial mass of the cloud. In addition, we assume an isothermal EOS, a cloud initial size $R_0 = 10$~pc and a mass transfer rate $\xi = 0$. The fragmentation rate $\phi(R)$ is computed for different cloud initial masses $M_0 = 10^3 - 10^4 - 10^5 - 10^6 ~\text{M}_\odot$, corresponding respectively to mean densities $\bar{\rho_0} \approx 10^{-22} - 10^{-21} - 10^{-20} - 10^{-19}~\text{g/cm}^{-3}$ with a cloud radius of $R_0/2$. The relative variations of the fragmentation rate are all similar even though more diffuse clouds have smaller fragmentation rates (Fig. \ref{fig:initial_condition}-(a)). 

In the subsonic regime associated with a thermal support, we notice that the fragment average masses and average densities are similar no matter the initial density (respectively Figs. \ref{fig:initial_condition}-(b) and -(c)). Indeed, as the spatial scale decreases the mass and density slowly converge towards the same numerical values. The asymptotic trend is analytically understandable within the framework of our model. On small scales, there is a convergence towards marginally stable structures as $\delta_c \xrightarrow[]{} 0$, which induces convergence of $\phi - \xi \xrightarrow[]{} 1$ (Eq.\ref{eq:subsonic_fragrate}). On the other hand, the difference $\phi - \xi$ corresponds to the local variation of average mass (Eq.\ref{eq:mass_size_relation}). Hence, if $\phi - \xi$ is constant, then

\begin{equation}
    \langle M \rangle ~= M_0 \left( \dfrac{R}{R_0} \right)^{\phi - \xi}
.\end{equation}

\noindent With $\phi - \xi = 1$ we have $\langle M \rangle \propto R$. This scaling relation is typical of the critical unstable Bonnor--Ebert isothermal gas spheres embedded in a pressurised medium \citep{Bonnor1956} {as} 

\begin{equation}
    M_{BE} = 2.4 R_{BE} \frac{C_s^2}{G}
,\end{equation}

\noindent where $R_{BE}$ is the critical {Bonnor--Ebert} radius of a spheroidal isothermal gas clump. In addition, under the same conditions, the average density of the fragments can be written as

\begin{equation}
    \langle \rho \rangle ~= \rho_0 \left( \dfrac{R}{R_0} \right)^{\phi - \xi - 3}
.\end{equation}

\noindent With $\phi - \xi = 1$ we have $\langle \rho \rangle \propto R^{-2}$. This scaling relation is what to be expected as a result from collapsing self-gravitating structures \citep{shu_self_similar_1977}. Both of these relations suggest that fragmentation is a mechanism tending to naturally produce marginally stable structures that collapse quasi-statically in order to form stars, regardless of the cloud initial stability (i.e. its initial density). Even though these structures have not actually reached a stage of marginal stability with $\delta_c \approx 0$, these asymptotic behaviours are reasonable approximations on a small scale $R < 0.01$~pc (Figs. \ref{fig:initial_condition}-(b) and -(c)).

Nevertheless, the similar convergence on both the mass and density numerical values suggests that fragmentation is also a self-regulated process. The more massive a parent is, the higher its fragmentation rate. As it fragments more, the average mass of the children is reduced accordingly to the value of $\phi - \xi$, as they need to share their parent mass. A less massive parent tends to fragment less, meaning its fragmentation rate is lower than that of a more massive parent. Thus, the average mass of each children decreases less rapidly. The two parents eventually form children with the same mass on average. As a consequence, these children fragments identically as they have the same fragmentation rate. 

\subsubsection{{The structure of the cloud}}

{In the supersonic regime associated with turbulent fragmentation, the average mass $\langle M \rangle$ evolves like a power law as a function of the spatial scale $R$ (Fig. \ref{fig:initial_condition}-(b)), reflecting the fractal structure of the cloud above 0.1~pc as a consequence of turbulence \citep{elmegreen1997, henne&falga, kritsuk2013}. However, the fragmentation rate, which traces the local slope as $\langle M \rangle \propto R^{\phi}$ considering $\xi = 0$, varies between $\phi(R) = 2 - 3$ for $R > 0.1$~pc (Fig. \ref{fig:initial_condition}-(a)). Hence, the apparent fractality of the cloud seen from the mass-size relation at large scales is an artifact of the low dynamical spatial range at which this fractality is considered. Nonetheless, this variation range is consistent with mass-size relation fractal index derived for example by \citealt{larson1981_relation} and \citealt{myers1983} ($M \propto R^2$), \citealt{traficante2018} ($M \propto R^2.38$) and \citealt{elmegreen_falgarone1996} {(between $M \propto R^{2.3 - 3.7}$)}. The variability of these slopes may result either from a difference in the mass transfer rates ($\xi$) within the clouds and/or as a difference in the cloud initial density $\bar{\rho}$ (Fig. \ref{fig:initial_condition}).}

\subsubsection{Characteristic scale as a stop for fragmentation}
\label{Sec:Natural stop for the fragmentation}

\begin{figure*}[!t]
    \centering
    \includegraphics[width=18cm]{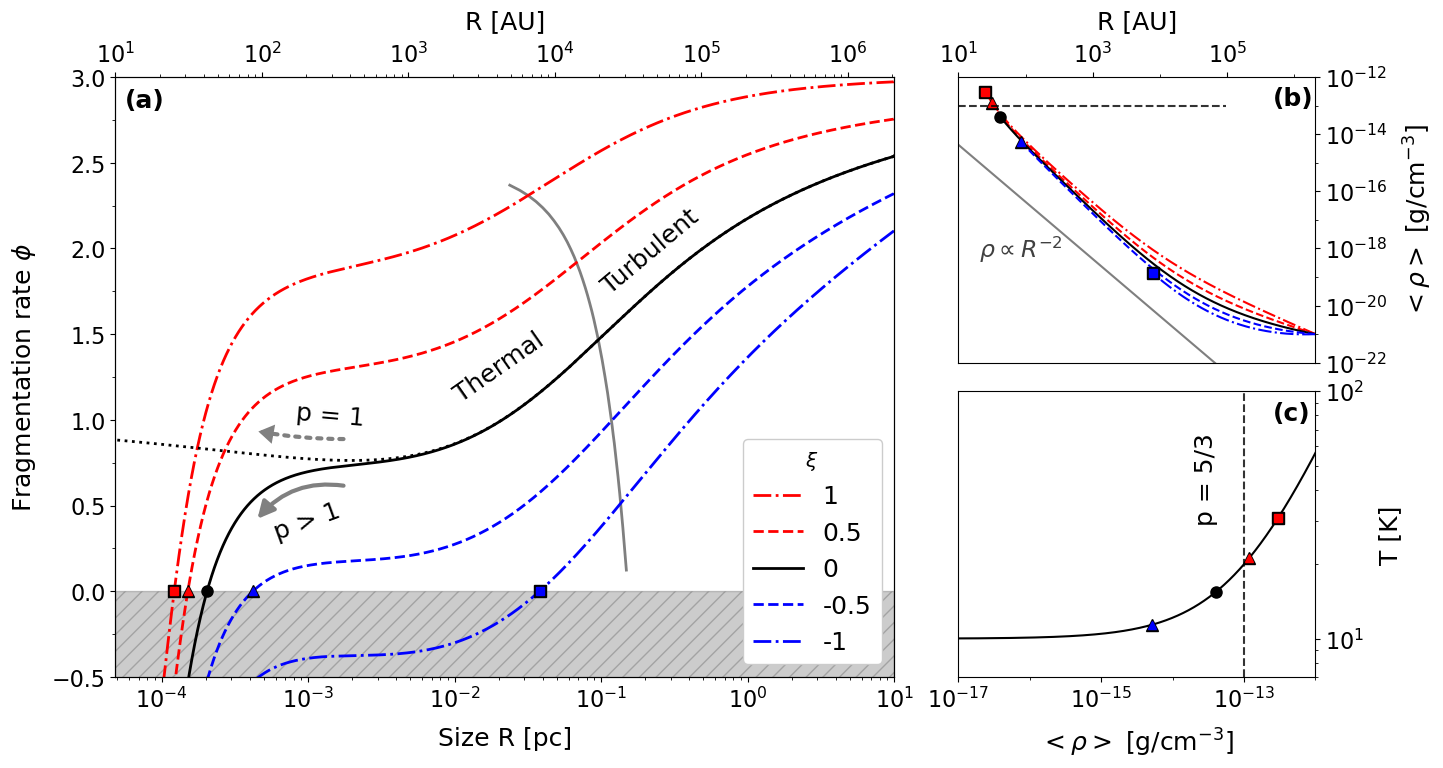}
    \caption{{Theoretical variations in the fragmentation rate $\phi$ (a), the mean fragment density $\langle \rho \rangle$ (b) as a function of the spatial scale $R$ for different mass transfer rates $\xi$. These profiles were computed using an adiabatic EOS (c), a cloud initial size $R_0 = 10$~pc with an initial density $\rho_0 = 10^{-21}$~g/cm$^{3}$. Red triangle, black dot, blue triangle, and blue square indicate the position of the last fragmenting structures for mass transfer rates $\xi = 1, 0.5, 0, -0.5, -1$, respectively. {(a)}: The black dotted line is associated with an isothermal EOS with a polytropic index $p = 1$. The transition between turbulent and thermal fragmentation is delimited by the  grey line. {(b)}: The black dashed line indicates the density of the first Larson core \citep{larson1969_fcore} when $p = 5/3$. {(c)}: Evolution of temperature $T$ as a function of fragments density.}}  
    \label{fig:phi_adiabatic}
\end{figure*}

To model the increasing opacity of the fragments as their density increases, we use the polytropic EOS given by Eq.\ref{eq:adia_EOS} with an adiabatic index $\gamma = 5/3$. We consider a cloud of size $R_0 = 10$~pc and mass $M_0 = 10^4~\text{M}_\odot$, {so the} initial mean density $\bar{\rho_0} \sim 10^{-21}$~g/cm$^3$. The fragmentation rate is computed for different mass transfer rates $\xi = \text{constants}$ varying between 1 and -1 (Fig. \ref{fig:phi_adiabatic}-(a)). 

For $R > 10^{-2}$~pc, the fragmentation rate follows isothermal behaviour (Fig. \ref{fig:isothermal_phi}-(a)) as the cloud is still under isothermal conditions with $T = 10$~K. However, unlike the case of pure isothermal EOS, $\phi(R)$ eventually becomes negative and goes into a {disrupted} regime as thermal pressure overcomes gravity. The spatial scales $R_\text{stop}$ at which $\phi(R) = 0$ so fragmentation ends are defined as the stopping scales. For each mass transfer rate we can attribute one different stopping scale that is typically at $R_\text{stop} \sim 40 - 100$~AU for $\xi \geqslant -0.5$ and $R_\text{stop} \sim 1$~kAU for $\xi = -1$. These characteristic scales of a few tens of AU represent the scales below which the fragmentation of a gaseous clump is prevented and {set} the initial conditions for the formation of the first Larson core \citep{Lee&H2018_adiafirstlarson}. However, at these scales the presence of a proto-planetary disk is expected. Such a disk structure is fundamentally  different from a spherical gas structure we have considered and have a different fragmentation process. The consequences of disk fragmentation are discussed more specifically in Sect.\ref{Sec:Disk fragmentation}.

The density of the last fragmenting structures formed of size $R_\text{stop}$ is represented in Fig. \ref{fig:phi_adiabatic}-(b) while their associated temperature is represented in Fig. \ref{fig:phi_adiabatic}-(c). For $\xi \geqslant 0.5$ the last fragmenting structures are adiabatic, setting the initial conditions for the formation of the first Larson core in these structures. The last fragments associated with $-0.5 < \xi < 0.5$ have started to heat up without reaching the adiabatic regime, {that is} the initial conditions for the first Larson core. It is necessary for these structures to get denser in order to host stars. Hence, they need to accrete from their environment and increase their mass transfer rate, otherwise any pre-stellar core of size $R < R_\text{stop}$ is expected to {disperse}. Then, if $\xi < -0.5$ the fragments are still isothermal and they are unlikely to form the first Larson core.

%
%

\section{Discussion}
\label{sec:section3}


\subsection{Observational constraints}
\label{Sec:Observational constraints}

\subsubsection{Number of YSOs in stellar groups}
\label{Sec:Number of YSOs in stellar groups}

Using the fragmentation rate solutions we computed within a gravo-turbulent framework in the case of an adabatic EOS (Sect.\ref{Sec:Natural stop for the fragmentation}), we can evaluate the average number of stars $\langle N_* \rangle_R$ located within spherical stellar group of size $R$ using Eq.\ref{eq:stellar_multiplicity}. As the fragmentation rate $\phi(R)$ cancels out for sizes $R_\text{stop}$, we assume below this point a one-to-one correspondence between the prestellar cores and young stars such that at $R_\text{stop}$ the number of stars $\langle N_* \rangle_{R_\text{stop}} = 1$.

The average number of stars $\langle N_* \rangle_{R}$ can be directly compared to the NESTs as defined by \cite{joncour2018nests} as spatially extended stellar structures, {that are local concentrations} of stars within the cloud. We show (Fig. \ref{fig:NEST_multiplicity}) that the spatial extension of NESTs of size $R > 20$~{kAU} is compatible with our model considering mass transfer rates $\xi \sim -0.75$ for an initial cloud density $\rho_0 = 10^{-20}$~g/cm$^{-3}$. In a star formation efficiency perspective, this means that a 0.1~pc clump produces the last fragments at scale $R_\text{stop} \approx 800$~AU with a $\langle \mathcal{E}(R_\text{stop}) \rangle \sim 9\%$ efficiency. For low clouds initial density $\rho_0 > 10^{-23}$~g/cm$^{-3}$, NESTs of size $R > 20$~kAU remain compatible with our model for $\xi < -0.25$, for average efficiencies $\langle \mathcal{E}(R_\text{stop}) \rangle ~< 20\%$ at stopping scales $R_\text{stop} \approx 40$~AU.

In addition, the number of stars within NESTs saturates around $\langle N_* \rangle_{R}^\text{NESTs} ~\sim 5$, between $R = 10$~kAU and $R = 20$~kAU. We can interprete this plateau with our model as the spatial properties of the smaller NESTs are compatible with higher mass transfer rates, suggesting that $\xi$ may not be a constant across the spatial scales but instead increases as the size of the fragments decreases. The plateau at $\langle N_* \rangle_{R}^\text{NESTs} \sim 5$ implies that $\phi = 0$ at least within the $R = 5 - 20$~kAU range of scales. Hence, the collapse should be monolithic at these scales. However, given the multiplicity of these NESTs, we expect that fragmentation resume on a spatial scales $R < 5$~kAU. The actual multiplicity of the fragments that could have generated these NESTs over this spatial range appears compatible with our model at scales $R > 20$~kAU considering constant mass transfer rate while the mass transfer rate should be increasing for $R < 20$~kAU suggesting a better formation efficiency or a more efficient accretion of mass at smaller scales, or both. Although, the NESTs multiplicity is well reproduced at all scales for diffuse cloud ($\rho_0 = 10^{-23}$~g/cm$^3$) considering a constant mass transfer rate $\xi = -0.25$.

\begin{figure}
    \centering
    \includegraphics[width=9cm]{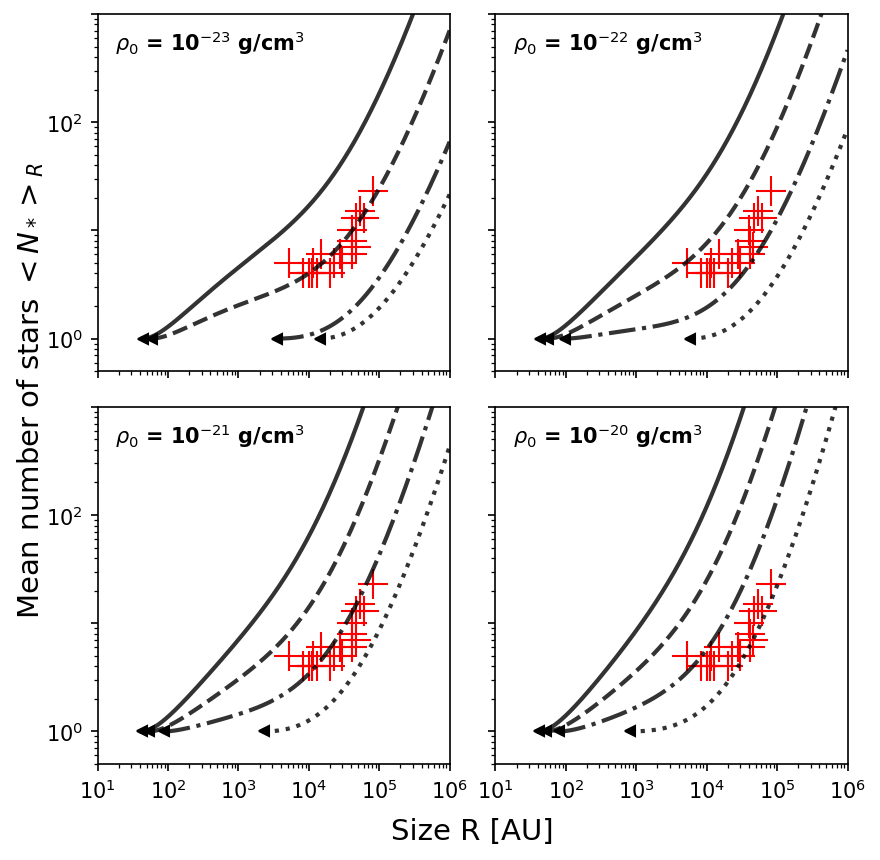}
    \caption{Average number of stars $\langle N_* \rangle_R$ located within a region of size $R$ for different initial cloud density $\rho_0$ as predicted by our model of fragmentation. The solid, dashed, dot-dashed, and dotted curve represent our model for different mass transfer rates $\xi = 0, -0.25, -0.5, -0.75$, respectively. The triangles mark the points below which gas clumps do not fragment anymore. The crosses represent the number of YSOs in NESTs \citep{joncour2018nests} as a function of their size, as observed in the Taurus cloud.}
    \label{fig:NEST_multiplicity}
\end{figure}

\subsubsection{Observational constraints on the fragmentation rate}
\label{Sec:Observational constraints on the fragmentation rate}

This model of fragmentation can be constrained using multi-scale analysis of molecular clouds. If one is able to count the number of fragments at different spatial scales, the fragmentation rate may be measured. Such a counting was performed using a combination of \textit{Spitzer} and \textit{Herschel} 70 - 160 - 250 - 350 and 500~$\mu$m images of NGC 2264 star-forming region \citep{thomasson2022}. In this work, we define a fractality coefficient $\mathcal{F}$ as the average number of fragments a single parent produces between two scales separated by a ratio $r = R_{l}/R_{l+1} = 2$. Using the definition of $\phi(R)$ given by Eq.\ref{eq:def_fragrate} we can define a two dimensional fragmentation rate $\phi_{2D}$ describing the variation of the average number of fragments $\langle N_{2D} \rangle$ observed in the two dimensional projection of a molecular cloud. Because the data sample the spatial scales into discrete levels of fragmentation (e.g. Fig.\ref{fig:fragmentation_network}), we cannot compute the `instantaneous' fragmentation rate at scale $R$ but instead the average fragmentation rate $\overline{\phi_{2D}}$ integrated over the range of scales $R$ that separate the largest fragment to the smallest one

\begin{equation*}
    \overline{\phi_{2D}} = ~ - ~~ \frac{\Delta \ln \langle N_{2D} \rangle}{\Delta \ln R}
.\end{equation*}

By definition of the fractality coefficient, there is a production of $\Delta \ln \langle N_{2D} \rangle = \ln \mathcal{F}$ fragments for every reduction of scales $\Delta \ln R = -\ln 2${, so}

\begin{equation}
    \overline{\phi_{2D}} = \dfrac{\ln \mathcal{F}}{\ln 2}
.\end{equation}

Subsequently we can compute the error on the fragmentation rate associated with the measure of $\mathcal{F}$ {as}

\begin{equation}
    \overline{\phi_{2D}} = \dfrac{\ln \mathcal{F}}{\ln 2} \pm \dfrac{\sigma_\mathcal{F}}{\mathcal{F} \ln 2}
.\end{equation}

\noindent With a measurement $\mathcal{F} = 1.45 \pm 0.12$ \citep{thomasson2022}, we obtain $\overline{\phi_{2D}} = 0.54 \pm 0.12$. To compute the equivalent three dimensional fragmentation rate $\overline{\phi_{3D}}$ we need to account for projection effects that may coincidentally align two structures on the line of sight and reduce the number of fragments we can detect. To obtain this three dimensional fragmentation rate, ellipsoidal geometries are randomly sampled and fragmented accordingly to a fragmentation rate $\overline{\phi_{3D}}$. The resulting ellipsoidal fragments are randomly placed inside their parents and the fragmentation process is repeated for at smaller scales. Between two levels of fragmentation, we used a scaling ratio $r = 1.5$, similar to the scaling ratios defined in \cite{thomasson2022} (more details in Appendix \ref{Appendix:comparisonNGC2264}). We then project the obtained fragments into a 2D plane and filter the aligned ones. We finally count the amount of fragments remaining after the projection in order to derive the fragmentation rate $\overline{\phi_{2D}}$. The error induced by the projection is given in Fig. \ref{fig:calibration_phi3D} in which we estimate a variation $f = \dfrac{\overline{\phi_{2D}} - \overline{\phi_{3D}}}{\overline{\phi_{3D}}} \approx -0.30 \pm 0.10$. From the relation

\begin{equation}
    \overline{\phi_{3D}} = \dfrac{\overline{\phi_{2D}}}{1 + f}
,\end{equation}

\noindent we can compute the uncertainty $\sigma_{\overline{\phi_{3D}}}$ with a quadratic error propagation

\begin{equation}
    \sigma_{\overline{\phi_{3D}}} = \sqrt{\left( \dfrac{\sigma_{\overline{\phi_{2D}}}}{1 + f}\right)^2 + \left( \dfrac{\overline{\phi_{2D}}}{(1 + f)^2} \sigma_f \right)^2}
.\end{equation}

With $\overline{\phi_{2D}} = 0.54$, $\sigma_{\overline{\phi_{2D}}} = 0.12$, $f = -0.3$ and $\sigma_f = 0.1$, we obtain the three dimensional average fragmentation rate

\begin{equation}
    \overline{\phi_{3D}} = 0.77 \pm 0.20
.\end{equation}

As this fragmentation rate has been computed within the $1.4 - 26$~kAU range, we can compute the mean fragmentation rate we obtain in the gravo-turbulent model (Sect.\ref{Sec:An analytical solution for the fragmentation rate}) for $\xi \in [-1 ; 1]$ with

\begin{equation}
    \overline{\phi_{3D}}_{\text{model}} = \dfrac{1}{\ln R_b - \ln R_a } \int_{\ln R_b}^{\ln R_a} \phi(R) d\ln R
    \label{eq:phi_moy_3D}
,\end{equation}

\noindent where $R_a = 1.4$~kAU and $R_b = 26$~kAU. We numerically computed the Eq.\ref{eq:phi_moy_3D} with the gravo-turbulent model with a $R_0 = 10$~pc cloud of initial mass $M_0 \approx 10^{4}$~M$_\odot$ to simulate NGC~2264 \citep{dahm2008} corresponding to initial cloud density $\rho_0 \approx 10^{-21}$~g/cm$^3$. We determine that the measured fragmentation rate is compatible with our theoretical value for mass injection rates $\xi \in [-0. 49 ; -0.15]$ (see Fig. \ref{fig:NGC2264solution}), meaning that between 7 and 40\% of the mass of a 0.1~pc gas structure is used to form its stars. In NGC~2264 hierarchical fragmentation is characterised by a loss of fragment mass with $\xi < 0$, for example via molecular jets as it has been observed in the fragmented hubs of the cloud \citep{maury_ngc_2009, cunningham_submillimeter_2016}. Since we define no proper physical model to describe $\xi$ we are not able to quantify the mass ejected per unit time for better comparison. Nonetheless, the range of scales ($R < 0.1$~pc) and the numerical value of the fragmentation rate ($\phi \sim 0.77$) suggests that the subfragmentation of NGC~2264 corresponds to an isothermal fragmentation (Fig. \ref{fig:phi_adiabatic}), dominated by gravity with a marginal contribution from turbulence. 

Furthermore, the loss of fragment mass with $\xi < 0$ evaluated in NGC~2264 is not incompatible with the NESTs properties from Taurus (see Sect.\ref{Sec:Number of YSOs in stellar groups}) despite the fact that NGC~2264 is a dense star-forming region while Taurus is a more diffuse cloud. We may hypothesise that $\xi < 0$ at scales $R > 10$~kAU is essentially the rule within molecular clouds regardless of their densities. As we show in Sect.\ref{sec:Turbulent vs thermal support}, these scales characterised by a negative mass transfer rate are also characterised by turbulent fragmentation. Hence, the mode of fragmentation (here turbulent) could set the stellar multiplicity rather than the cloud initial conditions.

\begin{figure}
    \centering
    \includegraphics[width=8cm]{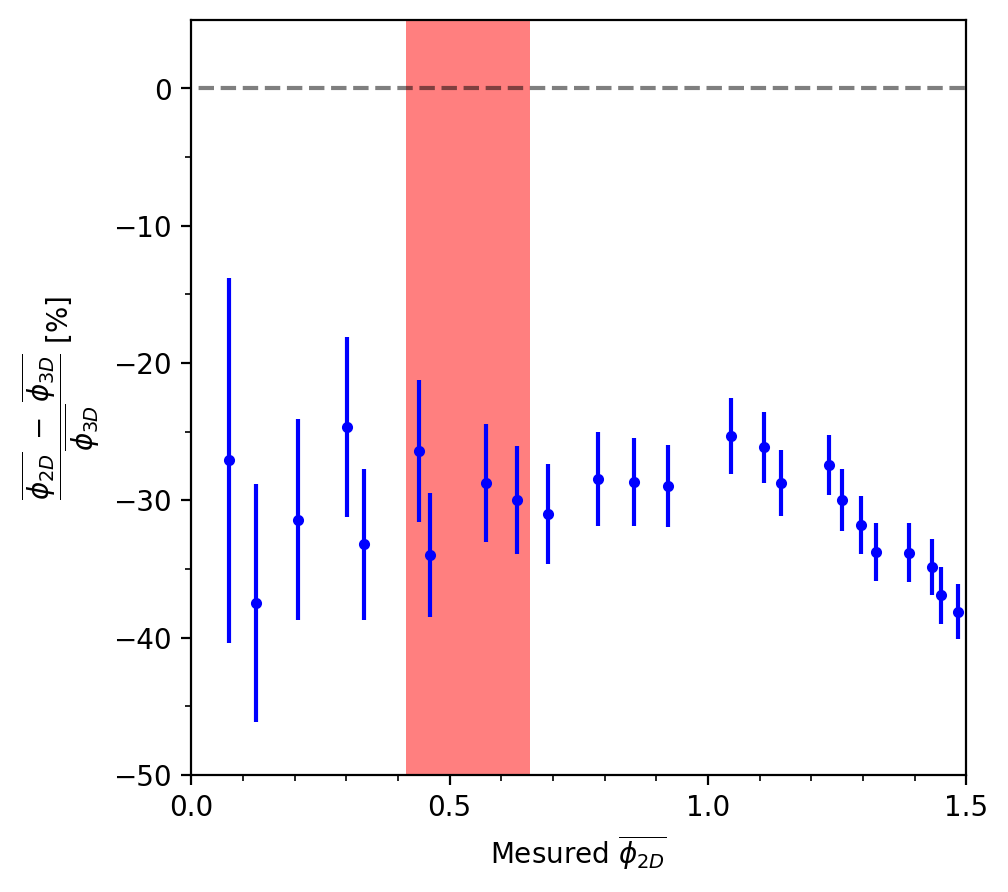}
    \caption{Relative error caused on the fragmentation rate by the projection of 3D ellipsoidal fragments onto a 2D plane. The 2D fragmentation rate $\overline{\phi_{2D}} = 0.54 \pm 0.12$ (red area) is associated with a relative error $\sim -30\% \pm 10$ with the 3D fragmentation rate $\overline{\phi_{3D}}$. The dashed grey line indicates a 0\% error.}
    \label{fig:calibration_phi3D}
\end{figure}

\begin{figure}
    \centering
    \includegraphics[width=8cm]{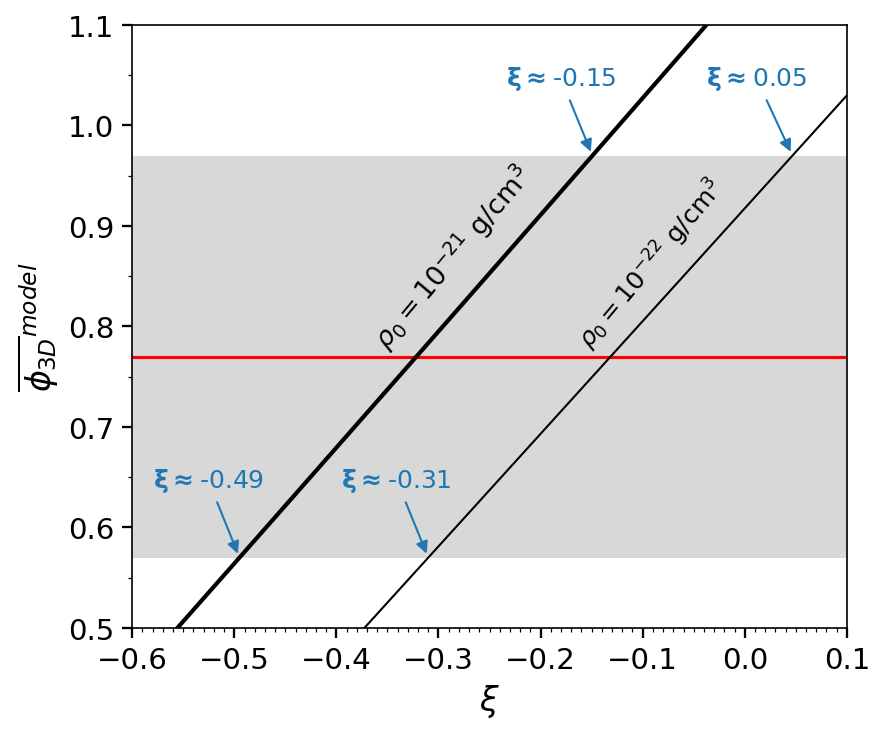}
    \caption{Theoretical average fragmentation rate $\overline{\phi_{3D}}_{model}$ integrated between $R_a = 1.4$~kAU and $R_b = 26$~kAU as a function of the injected mass rate $\xi$. The red horizontal line and grey uncertainty band corresponds to the fragmentation rate $\overline{\phi_{3D}} = 0.77 \pm 0.20$ measured in NGC~2264. The arrows indicate the $\xi$ boundaries between which our model is compatible with the measurement for different cloud initial densities $\rho_0$.}
    \label{fig:NGC2264solution}
\end{figure}


\subsection{Influence of a magnetic field}
\label{section:additional support}

Throughout our model we have considered a non-magnetised cloud. To qualitatively assess the local influence of a magnetic field, we consider that the presence of a magnetic field brings an additional support to counteract gravity, and therefore prevents fragmentation. To evaluate the variation induced by the magnetic field, or any mechanism that tends to stabilise a structure through additional support, we may add a constant velocity component $V_\text{add}$ to Eq.\ref{eq:delta_c}{, so}

\begin{equation}
    \delta_c = \ln \left[\frac{\pi^{5/3}}{6^{2/3}} ~ \frac{3 C_s^2 + V_\text{rms}^2 + V_\text{add}^2}{3G R^2 \bar{\rho}} \right]
.\end{equation}

To qualitatively estimate the additional support arising from the magnetic field, we introduce the Alfven velocity

\begin{equation}
    v_A = \sqrt{ \frac{B^2}{\mu_0 \rho} }
    \label{eq:alfven}
,\end{equation}

\noindent where $B$ is the magnetic field strength, $\mu_0 = 1.26.10^{-6}$~T~m/A is the {vacuum magnetic permittivity} and $\rho \sim 4.10^{-14}$~g/cm$^{-3}$ the fragment density at $R \sim 40 - 100$~AU. We computed the fragmentation rate $\phi(R)$ for $V_\text{add} < 500$~m/s (Fig. \ref{fig:influenceofsupport}) which corresponds to an equivalent magnetic fields $B < 30$~mG in order of magnitude (see e.g. \citealt{biersteker2019A, vlemmings2019}), assuming $v_A \sim V_\text{add}$. This additional support shifts the $R_\text{stop}$ from 40~AU towards larger scales, up to 100~AU as the fragmentation rate decreases. Nevertheless, the magnetic field in proto-planetary disks of tens of AU is typically about $B \sim 1$~mG \citep{brauer2017, lesur2023}. Considering this magnetic strength, the additional support is negligible as $v_A \sim 3$~m/s. We suppose here that the support $V_\text{add}$ is constant over the whole $R = 40$~AU - 10~pc range even though at scales $R > 0.1$~pc the magnetic field is negligible (B $\sim 10-30\mu$G, \citealt{crutcher2010, pattle2023}). Therefore, the variation of $R_\text{stop}$ caused by the magnetic-field should be taken as a maximum variation.

In addition to this support, accounting for the magnetic field in our model would break the isotropic assumption. As a consequence, the effective parental volume that can be filled by the children would be reduced. This is reflected in the {volume} ratio $V(R)/V_0 = (R/R_0)^{-3}$ in Eq.\ref{eq:generalnumber}. Under the influence of a magnetic field, fragmentation tends to be inhibited in the perpendicular direction of the magnetic field lines, reshaping the geometries of the structures. Because of this anisotropy, the size of fragments may vary more quickly in one direction compared to another (e.g. an ellipsoid that fragments along its length and the fragments have the same width of their parent). As a result, the power index -3 should decrease. As fragmentation is geometrically constrained by the effective parental volume, $\phi$ should decrease. Hence, even without introducing an additional support, the fragmentation rate can be reduced by any anisotropy.

\begin{figure}
    \centering
    \includegraphics[width=8cm]{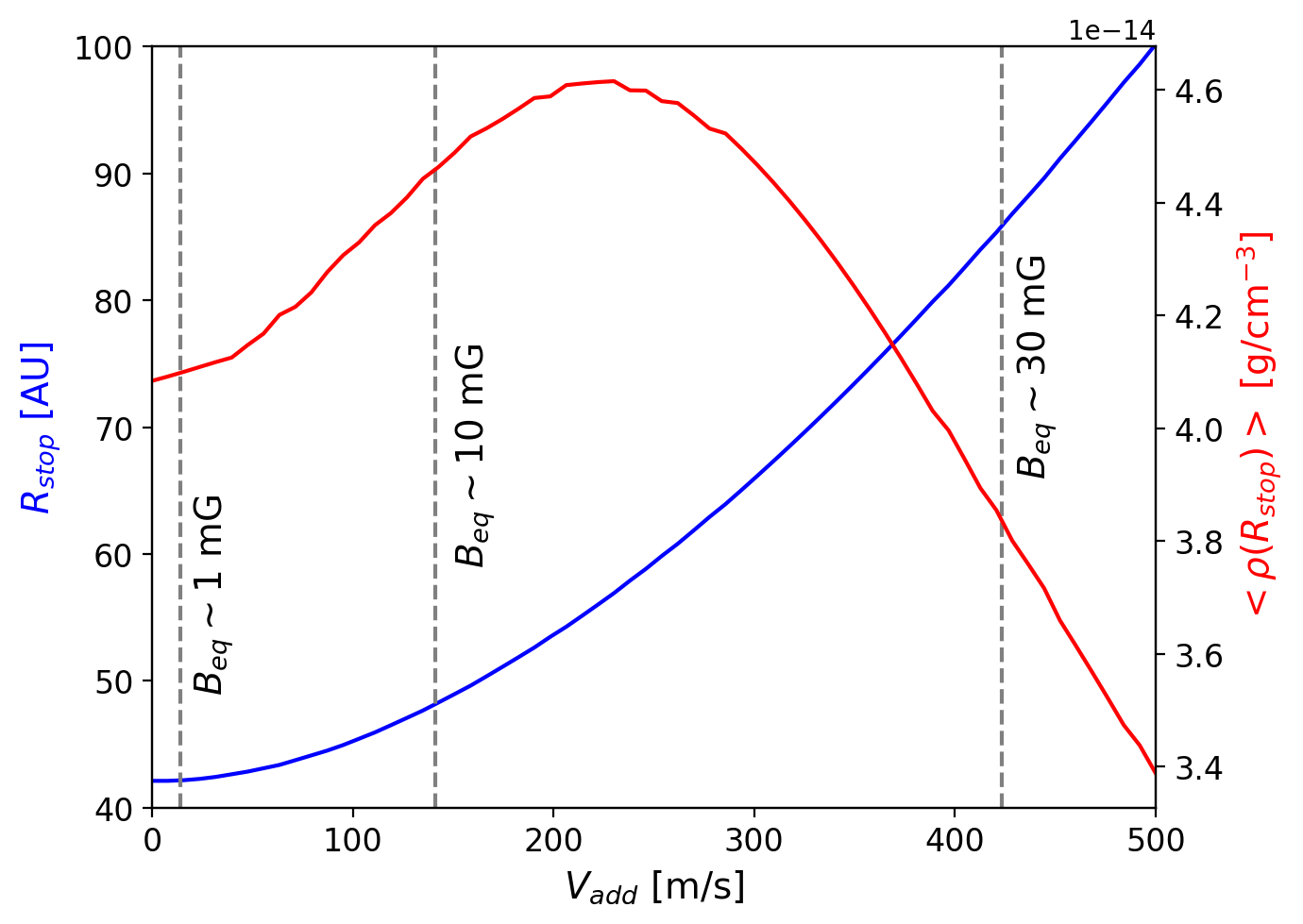}
    \caption{Influence of  additional support $V_\text{add}$ on both the fragmentation stopping scale $R_\text{stop}$ (blue) and the density of the last fragmenting structures $<\rho(R_\text{stop})>$ at this stopping scale (red). The {three vertical grey
    dashed lines} indicates the estimated equivalent magnetic field strengths from their Alfven velocities at $\rho \sim 4.10^{-14}$~g/cm$^{-3}$ using Eq.\ref{eq:alfven}.}
    \label{fig:influenceofsupport}
\end{figure}


\subsection{Disk fragmentation}
\label{Sec:Disk fragmentation}

Below the stopping scale $R_\text{stop} \sim 40 - 100$~AU, we expect the first Larson core to emerge and later the formation of a protoplanetary disk. Hence, the collapse criteria of a gaseous sphere $\delta_c$ we used to compute the fragmentation rate (Eq.\ref{eq:delta_c}) should not be valid anymore. Instead, one should consider the instability condition of a rotating disk that may be magnetised. The instability of such a disk may generate additional fragmentation events that result in the formation of binaries. Hence, to obtain the actual multiplicity of the close stellar systems and the mass of their stars it would be necessary to account for these fragmentation events.

At first approximation, we may assume that the prestellar core turns into a single star with a one-to-one correspondence. We can define the efficiency between the core and the star $\epsilon_\text{CS}$ representing the fraction of the prestellar core mass that effectively end up in the star it hosts. However, if multiple stars are formed because of disk fragmentation, {for example} a binary, this efficiency needs to be corrected by the mass ratio $q$ of the binary (see Sect.\ref{Sec:A stochastic and geometrical multi-scale model}). This mass partition would reduce the individual mass of the stars compared to a similar core that would only produce a single star. Hence, disk fragmentation would tend to produce less massive stars.

Furthermore, the systematic formation of a binary would systematically multiply the number of stars produced within gaseous fragments by a factor of $2$. Thus, the spatial and multiplicity properties of the NESTs would become compatible for mass transfer rates $\xi < -0.75$ for scales $R > 0.1$~pc suggesting that supersonic turbulence produces fragments with low mass efficiencies. As an order of magnitude, a $R_0 = 10$~pc cloud produces $R = 0.1$~pc fragments with an efficiency $\langle \mathcal{E} \rangle = (0.1/10)^{-\xi} \sim 3\%$ with $\xi = -0.75$. Under such circumstances the fragments {do} not reach their adiabatic phase as their density is not high enough, assuming $\xi = -0.75$. This suggests that the mass injection rate $\xi$ increases for $R < 0.1$~pc in order for the fragments to reach their adiabatic phase and effectively form a star.


\subsection{The mass of the stars}

In Sect.\ref{Sec:Self-regulation of the fragmentation} we have shown that fragmentation is self-regulated at small scales, suggesting that the final properties of stars are marginally determined by the initial conditions of the cloud. Provided that this self-regulation is sufficiently fast to occur, the young stars or even pre-stellar cores should possess a characteristic average mass independant of their cloud. {Under adiabatic conditions, hierarchical fragmentation leads to fragments of size $R_\text{stop}$ with characteristic mass $M \approx 4.10^{-3}$~M$_\odot$. These low mass structures are consistent with fragments built under adiabatic conditions when they reach their opacity limit \citep{whitworth_boyd2005, Lee&H2018_adiafirstlarson}. Nonetheless, the mass of the first Larson core that supposedly forms inside these fragments is about $10^{-2}$~M$_\odot$ \citep{larson1969_fcore, Lee&H2018_adiafirstlarson}, while the characteristic stellar mass is $0.1$~M$_\odot$, corresponding to the `peak' of the IMF. This characteristic stellar mass can emerge within the framework of a single step fragmentation process (i.e. without hierarchy) in which filaments can produce fragments of mass $0.1 - 10$~M$_\odot$ \citep{larson1985_critmass}. Similarily, the thermal Jeans mass of $\sim 0.1$~pc clumps is about $0.1 - 1$~M$_\odot$ assuming no hierarchical substructures \citep{larson2005jeans, alves2007}. Hence, the introduction of a hierarchical process inherently build less massive fragments.}



{Using our model and the observational constraints we derived, we propose a scenario that may reconcile the mass discrepancies between the stars and the pre-stellar structures.} For scales $R > 0.1$~pc, we extract a mass transfer rate $\xi < 0$ (Sect.\ref{Sec:Observational constraints}). At these scales, turbulence is in a supersonic regime and the fragments are not completely bound because of turbulent dissipation. As a result, they {lose} a fraction of their mass during their formation {so} $\xi < 0$. On scales of $R < 0.1$~pc, turbulence enters a subsonic regime. The fragments at these scales are more bound than at larger scales so they conserve their mass better. This mechanism would result in a progressive increase of the mass transfer rate $\xi(R)$ as $R$ decreases {and} as the turbulent support vanishes{,} while these fragments {finally} accrete their environment to subsequently form stars.

{Alternatively, hierarchical fragmentation may have transient phases in which fragments collapse monolithically in a range of spatial scales before subfragmenting hierarchically again. This solution would produce more massive fragments as fewer fragments share the mass reservoir. However, this solution is speculative and the physical mechanism that would slow down the hierarchical process remains unkown.}

%
%

\section{Conclusion}
\label{sec:section4}

\subsection{{Summary}}

We introduced a probabilistic multi-scale fragmentation model to describe the multi-scale structure of molecular clouds down to star formation. With this model we aim to predict
the stellar multiplicity of the systems formed through hierarchical fragmentation as well as the mass of the fragments produced. This model connects the discrete nature of fragmentation, which results in the formation of countable stars, with the continuous,  multi-scale structure of the molecular clouds from which young stars inherit their mass. The discrete nature of fragmentation process is represented as a network organised into different spatial scales populated by fragments interconnected according to their genealogical affiliation. This architecture provides a framework to perform a multi-scale analysis that can be compared to the continous model of fragmentation. Our model introduces two continuous parameters that determine the multiplicity and mass of fragments at any spatial scale $R$: 

\begin{enumerate}
    \item the fragmentation rate $\phi(R)$, which  controls the amount of fragments produced at each scale and the multiplicity of young stars in stellar systems;
    \item the mass transfer rate $\xi(R)$, characterises the efficiency of fragment  formation.
\end{enumerate}

The fragmentation rate and mass transfer rate were used to compute the average number of fragments produced, their average mass, their average mass density, and their average stellar density. We also associated three fragmentation modes considering the numerical values of the fragmentation rate $\phi$ and the mass transfer rate $\xi$:

\begin{enumerate}
    \item {a dispersal} mode if $\xi < -3$, or by definition if $\phi > 0$;
    \item {an effective} monolithic collapse mode if $\phi \geqslant \xi + 3 \geqslant 0$, or if $\phi = 0$;
    \item {a} hierarchical fragmentation mode if $0 < \phi < \xi + 3$.
\end{enumerate}

Then we implemented a gravo-turbulent fragmentation in our model to analytically compute the fragmentation rate $\phi${,} evaluate its variation along the spatial scales, {and compare it with observational measurement}. The mass transfer rate $\xi$ was linked to the mass budget of fragment formation that represents the balance between accretion processes \citep{bonnell2001, longmore_formation_2014}, which    add mass to the fragments, and outflow processes \citep{krumholtz2012, clark2021} {or turbulence \citep{camacho2016}}, which deplete the fragments of their mass. Although this balance was not   explicitly modelled, it is necessary if one needs to quantitatively constrain the mass transfer rate $\xi$ with respect to any mechanism that locally modifies the mass contained within the fragments. {We highlighted three major results regarding the cloud structure, the fragmentation property, and the resulting star formation scenario.}

\paragraph{{Cloud structure:}}
{
We find that molecular clouds do not systematically exhibit fractal structure as the fragmentation rate is not scale-free in the general case. We highlight two characteristic spatial scales affecting fragmentation. The first   is the typical sonic scale $R \approx 0.1$~pc, from which fragmentation shifts from a turbulent regime to a thermal regime. The second scale $R_\text{stop} \sim 40$~AU marks the point beyond which a structure becomes optically opaque to its own radiation field, preventing further fragmentation from occurring, while allowing the first Larson core to form. Under supersonic turbulence the cloud may exhibit fractal properties at large scales. Similarily in the subsonic regime, scale-free Jeans-like fragmentation can be retrieved for marginally unstable fragments with an isothermal EOS.
}

\paragraph{{Fragmentation inherent property:}}
{
This gravo-turbulent model of fragmentation also revealed that fragmentation is a self-regulated process. At similar spatial scales, comparatively denser structures tend to fragment more, producing more diffuse fragments that subsequently subfragment less and form denser fragments. As a consequence, fragmentation progressively loses the information about the initial conditions of the cloud. At small scales, fragments eventually reach marginal instability as mass evolves linearly with their spatial scale $M \propto R$, characteristic of marginally unstable isothermal Bonnor--Ebert structures. 
}

\paragraph{{Star formation scenario:}}
{
In our model, hierarchical fragmentation produces pre-stellar structures of typical mass $M \sim 10^{-3}~$M$_\odot$, for which mass accretion is necessary to produce stars of characteristic mass $M \sim 10^{-1}~$M$_\odot$. On the other hand, the comparison of our model with the Taurus NESTs' multiplicity and size \citep{joncour2018nests} indirectly suggests that fragments are produced more efficiently at low scales, which implies that the mass transfer rate $\xi(R) < 0$   increases as the spatial scale $R$ decreases. These measurements are compatible with a sequential star formation scenario: in the early stages of fragmentation at large scales, dense structures are not formed efficiently as {random turbulent motion could unbind fluid particles from the fragments} material ($\xi < 0$). Then, as thermal instability becomes the dominant driver for fragmentation, turbulence {mass dispersal} is progressively reduced ($\xi$ increases). Finally, pre-stellar structures accrete their surrounding material ($\xi > 0$) at small scales without fragmenting so that stars can form.
}


\subsection{{Potential of this model}}


The main purpose of our model is to test hierarchical fragmentation as the mechanism setting the discrete properties of stars with respect to their diffuse environment at larger scale. Under  mass-loss conditions for the fragments formation (i.e. $\xi < 0$), the fragmentation cascade can explain the origin of NESTs \citep{joncour2018nests} and the subfragmentation of gas clumps observed in NGC~2264 \citep{thomasson2022}. However, further investigation is needed in order to fully map the transition between the dense structures located within molecular clouds and the young stars they form through fragmentation. In particular, we did not quantify how stellar IMF may rise from the fragmentation of cores composing a CMF \citep[Thomasson in prep.]{larson1973_fragmentation, elmegreen1983, GuszejnovHopkins2015}, nor did we assess how these stellar objects are dynamically bound in order to form stable multiple systems. {In addition, the model we implement focuses  on the isotropic fragmentation of roundish gas clumps, but our framework can be adapted and improved to investigate the fragmentation of alternative geometries such as filaments, galactic disks, or proto-planetary disks. To describe different objects, different turbulent or magnetic prescriptions should be implemented so the model can be reprocessed in order to derive different fragmentation rate functions.}

In order to obtain better constraints on the fragmentation rate, {multi-scale analysis should be performed on a wide range of spatial scales, for example within different molecular cloud environments}. {We can also challenge the model's theoretical prediction, given a set of implemented physical processes,} using hydrodynamical or magneto-hydrodynamical numerical simulations of molecular clouds down to core scale. {These simulated data would provide useful theoretical constraints regarding the fragmentation rate. As these simulations describe the evolution of a cloud it could be possible to evaluate the temporal dependency of both the fragmentation rate and the mass transfer rate}.

Although some observations have highlighted a correlation between the mass efficiency and the column density of dense clumps \citep{louvet_w43-mm1_2014}, no information is provided concerning the multi-scale behaviour of the mass efficiency. As the fragmentation rate, numerical simulations could be employed in order to measure the mass transfer rate. Hence, we could build a semi-empirical law to constrain our fragmentation model and have a mean for comparison with indirect observations. In a more general perspective,   measurements of the fragmentation rate and mass transfer rate could both be used as a proxy to {directly} compare, under the same theoretical basis, the multi-scale properties of the structures observed in real molecular clouds, {as we did in this work with NGC~2264}, with those that emerge from the {numerical} simulations.

\begin{acknowledgements} 
{The authors wish to acknowledge the anonymous referee whose useful comments helped us to clarify this work.} BT carried out this project under equal supervision of IJ and EM who both provided comments and suggestions on this manuscript. FL, FM, MG and TN provided helpful discussion and comments on the paper. We thank Patrick Hennebelle for in-depth discussions which greatly improved the quality of this work. BT has received financial support from the European Research Council synergy grant “ECOGAL” (Grant 855130), the “StarFormMapper” project funded by the European Union’s Horizon 2020 Research and Innovation Action programme (Grant 687528) and from the French Agence Nationale de la Recherche (ANR) through the project “COSMHIC” (ANR-20-CE31-0009).
\end{acknowledgements}

%
%

\bibliographystyle{aa} 
\bibliography{biblio}

%
%
\appendix


\section{Calculation of the fragmentation rate}
\label{Appendix:computefrag}

This Appendix aims to detail the formula for the fragmentation rate $\phi(R, \bar{\rho})$ given in Sect.\ref{Sect:Procedure to compute the fragmentation rate}. We recall here the Eq.\ref{eq:generalphi} which determines the fragmentation rate as a function of the probability {$\mathcal{P}(\delta \geqslant \delta_c)$} that a density fluctuation $\delta$ induces a local collapse with respect to a threshold fluctuation $\delta_c$:

\begin{equation}
 \phi(R) = 3 - \frac{d \ln \mathcal{P}(\delta \geqslant \delta_c)}{d \ln R}
 \label{annBeqorigin}
,\end{equation}

\noindent with

\begin{equation}
     \mathcal{P}(\delta \geqslant \delta_c) = \frac{1}{2} \left( 1 - \text{erf} \left[ \frac{\delta_c + \frac{\sigma^2}{2}}{\sigma \sqrt{2}} \right] \right)
.\end{equation}

Our aim is to compute $\dfrac{d \ln \mathcal{P}}{d \ln R}$. Let $u = \dfrac{\delta_c + \frac{\sigma^2}{2}}{\sigma \sqrt{2}}$ such that

\begin{equation}
    \frac{d \ln \mathcal{P}}{d \ln R} = - \frac{1}{2\mathcal{P}} \frac{d \text{erf}(u)}{d u} \frac{d u}{d\ln R}
.\end{equation}

As $\dfrac{d \text{erf}(u)}{d u} = \dfrac{2}{\sqrt{\pi}} \exp [-u^2]$ we obtain

\begin{equation}
    \frac{d \ln \mathcal{P}}{d \ln R} = - \frac{1}{\mathcal{P}} \frac{\exp [-u^2]}{\sqrt{\pi}} \frac{d u}{d\ln R} 
.\end{equation}

Expanding the derivative $\dfrac{d u}{d\ln R}$ we finally obtain

\begin{equation}
    \dfrac{d u}{d\ln R} = \dfrac{1}{\sigma \sqrt{2}} \left[ \frac{d \delta_c}{d \ln R} - \sqrt{2}~v \frac{d \sigma}{d \ln R}\right]
,\end{equation}

\noindent where we let $v = \dfrac{\delta_c - \frac{\sigma^2}{2}}{\sigma \sqrt{2}}$

Let $A_\text{p}\bigg(\delta_c \big( R\big) ; \sigma \big( R\big) \bigg) = - \dfrac{\exp[-u^2]}{\mathcal{P}\sigma \sqrt{2\pi}}$ so we can write

\begin{equation}
    \frac{d \ln \mathcal{P}}{d \ln R} = A_\text{p} \left[ \dfrac{d \delta_c}{d\ln R} - \sqrt{2}~v \dfrac{d \sigma}{d \ln R} \right]
    \label{eq:app:Pdeltasigma}
.\end{equation}

\subsection{Critical threshold}

First we compute the derivative of $\delta_c$ appearing in the expression of $\dfrac{d \ln \mathcal{P}}{d \ln R}$ of Eq.\ref{eq:app:Pdeltasigma}. We use the threshold determined by \cite{H&C2008} that consider both thermal and turbulent support:

\begin{equation}
    \delta_c = \ln \left[\frac{\pi^{5/3}}{6^{2/3}} ~ \frac{3 C_s^2 + V_\text{rms}^2}{3G R^2 \bar{\rho}} \right]
    \label{eq:annexe_phi:delta_c}
,\end{equation}

\noindent where $G = 6.67 \times 10^{-11}~\text{m}^3\text{kg}^{-1}\text{s}^{-2}$ is the gravitational constant, $C_s$ is the sound velocity, $\bar{\rho}$ the parent mean density and $V_\text{rms}$ the turbulent velocity. 

We can compute the derivative of $\dfrac{d \delta_c}{d\ln R}$ appearing in Eq.\ref{eq:app:Pdeltasigma} that sets the fragmentation rate {as}

\begin{equation}
    \frac{d \delta_c}{d \ln R} =
    \dfrac{ 3\dfrac{dC_s^2}{d \ln R} + \dfrac{dV_\text{rms}^2}{d \ln R} - \dfrac{3C_s^2 + V_\text{rms}^2}{3GR^2\bar{\rho}} \dfrac{d (3GR^2\bar{\rho})}{d \ln R} }{ 3C_s^2 + V_\text{rms}^2}
.\end{equation}
    
With

\begin{equation}
    \frac{1}{3GR^2\bar{\rho}} \frac{d (3GR^2\bar{\rho})}{d \ln R} = \frac{d \ln (3GR^2\bar{\rho})}{d \ln R} = \phi - \xi - 1
,\end{equation}

\noindent we obtain

\begin{equation}
    \frac{d \delta_c}{d \ln R} = \dfrac{3\dfrac{dC_s^2}{d \ln R} + \dfrac{dV_\text{rms}^2}{d \ln R}}{3C_s^2 + V_\text{rms}^2} + \xi - \phi + 1
    \label{eq:app:ddeltac_raw}
.\end{equation}

\subsubsection{Derivative of $C_s^2$}

At temperature $T$ the sound velocity can be expressed as

\begin{equation}
    C_s^2 = \frac{k_B T}{\mu m_H}
.\end{equation}

\noindent where $k_B = 1.38 \times 10^{-23}~\text{J/K}$, $\mu = 2.33$ et $m_H = 1.67 \times 10^{-27}~$kg respectively the Boltzmann constant, mean molecular weight and hydrogen mass. By computing

\begin{equation}
    \frac{1}{C_s^2} \frac{d C_s^2}{d \ln R} = \frac{d \ln C_s^2}{d \ln R} 
,\end{equation}

\noindent and considering the Eq.\ref{eq:rho_size_relation} that express the log-derivation of density, we obtain

\begin{equation}
    \frac{d C_s^2}{d \ln R} = C_s^2 ~ \frac{d \ln T}{d \ln \rho} ~ \bigg(\phi - \xi - 3\bigg)
    \label{eq:app:dCs}
.\end{equation}

\subsubsection{Derivative of $V_\text{rms}^2$}

The quadratic velocity resulting from Larson's laws \citep{larson1981_relation} is expressed as

\begin{equation}
    V_\text{rms}^2 = V_0^2 \left( \frac{R}{\text{1 pc}}\right)^{2\eta}
.\end{equation}

\noindent where $V_0 = 1000~\text{m/s}$ and $\eta = 0.38$ represents respectively the velocity dispersion of a region of size $R = 1$~pc and the power index characterising the turbulent cascade across the spatial scales. By computing

\begin{equation}
\frac{1}{V_\text{rms}^2} \frac{d V_\text{rms}^2}{d \ln R} = \frac{d \ln V_\text{rms}^2}{d \ln R}
,\end{equation}
    
\noindent we obtain

\begin{equation}
    \frac{d V_\text{rms}^2}{d \ln R} = 2\eta V_\text{rms}^2
    \label{eq:app:dVrms}
.\end{equation}

\subsubsection{Derivative of $\delta_c$}

Injecting both the derivatives of sound and turbulent velocities Eqs.\ref{eq:app:dCs} and \ref{eq:app:dVrms} within the expression of $\frac{d \delta_c}{d \ln R}$ Eq.\ref{eq:app:ddeltac_raw}, we finally obtain

\begin{equation}
    \frac{d \delta_c}{d \ln R} = \bigg( \xi - \phi + 3 \bigg) \left[ 1 - \hat{C_s}^2 \frac{d \ln T}{d \ln \rho} \right] - 2 \left[ 1 - \eta \hat{V}_\text{rms}^2 \right]
,\end{equation}

\noindent where we defined the reduced variables $\hat{C_s}^2$ and $\hat{V}_\text{rms}^2$:

\begin{equation}
    \begin{cases}
    \hat{C_s}^2 = \dfrac{3C_s^2}{3C_s^2 + V_\text{rms}^2} \\[15pt]
    \hat{V}_\text{rms}^2 = \dfrac{V_\text{rms}^2}{3C_s^2 + V_\text{rms}^2} = 1 - \hat{C_s}^2
    \end{cases}
.\end{equation}
    
\subsection{Variance of density fluctuations}

Next we compute the derivative of $\sigma(R)$ appearing in the expression of $\frac{d \ln \mathcal{P}}{d \ln R}$ of Eq.\ref{eq:app:Pdeltasigma} with

\begin{equation}
    \sigma^2 = \ln \left[ 1 + b^2 \mathcal{M}^2 \right]
,\end{equation}

\noindent where $b^2 \approx 0.25$. We compute

\begin{equation}
    \frac{d \sigma^2}{d \ln R} = \frac{d \mathcal{M}^2}{d \ln R} \dfrac{b^2}{1 + b^2 \mathcal{M}^2}
    \label{eq:app:sigma2_raw}
.\end{equation}

\subsubsection{Derivative of $\mathcal{M}^2$}

The Mach number $\mathcal{M}$ is defined as the ratio between the turbulent velocity to the sound velocity

\begin{equation}
    \mathcal{M} = \frac{V_\text{rms}}{C_s}
.\end{equation}

By computing $\frac{1}{\mathcal{M}^2} \frac{d \mathcal{M}^2}{d \ln R}$ and substituing both the derivatives of sound and turbulent velocities by Eqs.\ref{eq:app:dCs} and \ref{eq:app:dVrms} obtain

\begin{equation}
    \frac{d \mathcal{M}^2}{d \ln R} = \mathcal{M}^2 \left[ 2 \eta - \frac{d \ln T}{d \ln \rho} \bigg(\phi - \xi - 3\bigg) \right]
    \label{eq:app:mach}
.\end{equation}

\subsubsection{Derivative of $\sigma^2$}

Injecting the derivatives of the Mach number Eq.\ref{eq:app:mach} within the expression of $\frac{d \sigma^2}{d \ln R}$ Eq.\ref{eq:app:sigma2_raw}, we finally obtain

\begin{equation}
    \frac{d \sigma}{d \ln R} = \frac{\hat{\mathcal{M}}^2}{2\sigma} \left[ 2 \eta + \frac{d \ln T}{d \ln \rho} \bigg(\xi - \phi + 3\bigg) \right]
,\end{equation}

\noindent where we defined the reduced variables $\hat{\mathcal{M}}^2$:

\begin{equation}
    \hat{\mathcal{M}}^2 = \dfrac{b^2 \mathcal{M}^2}{1 + b^2 \mathcal{M}^2}
.\end{equation}

\subsection{The fragmentation rate}

We now have the following equations:

\begin{equation}
    \begin{cases}
     \dfrac{d \delta_c}{d \ln R} = \bigg( \xi - \phi + 3 \bigg) \left[ 1 - \hat{C_s}^2 \dfrac{d \ln T}{d \ln \rho} \right] - 2 \left[ 1 - \eta \hat{V}_\text{rms}^2 \right]\\[15pt]
        \dfrac{d \sigma}{d \ln R} = \dfrac{\hat{\mathcal{M}}^2}{2\sigma} \left[ 2 \eta + \dfrac{d \ln T}{d \ln \rho} \bigg(\xi - \phi + 3\bigg) \right] \\[15pt]
        \dfrac{d \ln \mathcal{P}}{d \ln R} = A_\text{p} \left[ \dfrac{d \delta_c}{d\ln R} - \sqrt{2}~v \dfrac{d \sigma}{d \ln R} \right] \\[15pt]
        \phi(R) = 3 - \dfrac{d \ln \mathcal{P}}{d \ln R}
    \end{cases}
    \label{eq:app:setphi}
.\end{equation}

We showed that both the critcal threshold $\delta_c$ and the standard deviation $\sigma$ depend on the fragmentation rate $\phi$ but also on the mass transfer rate $\xi$. To obtain the actual function that describes the fragmentation rate at a given $R$, one needs to isolate $\phi(R)$. After refactoring the set of equations (Eqs. \ref{eq:app:setphi}) and isolating the fragmentation rate we finally find

\begin{equation}
    \phi(R) = \frac{3}{1 - A_\text{th}} - \dfrac{A_\text{th}\big( \xi + 3 \big)}{1 - A_\text{th}} + \frac{2A_\text{turb}}{1 - A_\text{th}}
    \label{eq:app:fragrate}
,\end{equation}

\noindent where we let:

\begin{equation}
    \begin{cases}
        A_\text{p} = - \dfrac{\exp[-u^2]}{\mathcal{P}\sigma \sqrt{2\pi}} \\ \\
        A_\text{th} = A_\text{p} \left[ 1 - \dfrac{d \ln T}{d \ln \rho} \bigg( \hat{C_s}^2 + \dfrac{v \hat{\mathcal{M}}^2}{\sqrt{2} \sigma} \bigg) \right] \\ \\
        A_\text{turb} = A_\text{p} \left[1 -  \eta \bigg( \hat{V}_\text{rms}^2 - \dfrac{v \hat{\mathcal{M}}^2}{\sqrt{2} \sigma}\bigg) \right]
    \end{cases}
    \label{eq:ann:athpturb}
,\end{equation}

\noindent and:

\begin{equation}
    \begin{cases}
        u = \dfrac{\delta_c + \frac{\sigma^2}{2}}{\sigma \sqrt{2}} \\ \\
        v = \dfrac{\delta_c - \frac{\sigma^2}{2}}{\sigma \sqrt{2}}
    \end{cases}
    \label{AnnB:eq:Lesuetv}
.\end{equation}


\section{Asymptotic limits}
\label{Appendix:asymptotic}

\subsection{Subsonic limit}

First we assume that $V_\text{rms} \ll C_s$ such that turbulent support is negligible. In this limit, the Mach number $\mathcal{M}^2$ goes to 0. By using the expansion $\ln(1 + x) \approx x$, we also compute the asymptotic limit of density fluctuations variance $\sigma^2$:

\begin{equation}
    \sigma^2 \approx b^2 \left( \frac{V_\text{rms}}{C_s}\right)^2 
,\end{equation}

\noindent which goes to 0 in the subsonic limit. We can also compute in this limit the following asymptotic relations:

\begin{equation}
\begin{cases}
    \lim_{\sigma \to 0 } u = \dfrac{\delta_c}{\sigma \sqrt{2}} \\[15pt]
    \lim_{\sigma \to 0 } v = \dfrac{\delta_c}{\sigma \sqrt{2}} \\[15pt]
    \lim_{\mathcal{M} \to 0 } \hat{C_s}^2 = 1 \\[15pt]
    \lim_{\mathcal{M} \to 0 } \hat{V}_\text{rms}^2 = 0 \\[15pt]
    \lim_{\mathcal{M} \to 0 } \hat{\mathcal{M}}^2 \approx \sigma^2 \approx b^2 \left(\dfrac{V_\text{rms}}{C_s}\right)^2
\end{cases}
.\end{equation}

We may also compute the asymptotic limit of the quantity $\dfrac{v \hat{\mathcal{M}}^2}{\sqrt{2} \sigma}$ involved within the coefficients $A_\text{th}$ and $A_\text{turb}$ as it will become useful in the following:

\begin{equation}
    \lim_{\sigma \to 0 } \dfrac{v \hat{\mathcal{M}}^2}{\sqrt{2} \sigma} = \frac{\delta_c}{2}
.\end{equation}

Next we can compute the asymptotic limit of $A_\text{p}$ as given in Eq.\ref{eq:ann:athpturb}. In the subsonic limit the variable $u$ goes to $+\infty$ if $\delta_c > 0$ or goes to $-\infty$ if $\delta_c < 0$. Assume first that $\delta_c > 0$, we computed the case $\delta_c < 0$ in Appendix \ref{app:deltacnegative}. With $\delta_c > 0$

\begin{equation}
    \lim_{u \to +\infty } \mathcal{P}(\delta \geqslant \delta_c) = \frac{1}{2} \bigg( 1 - \lim_{u \to +\infty } \text{erf} \big(u\big) \bigg)
.\end{equation}

The expansion for $u \xrightarrow{} +\infty$ of the error function is

\begin{equation}
    \text{erf}(u) \approx 1 - \dfrac{\exp[-u^2]}{u\sqrt{\pi}}
.\end{equation}

In that case we can write $\mathcal{P}(\delta \geqslant \delta_c)$ as

\begin{equation}
    \lim_{u \to +\infty} \mathcal{P}(\delta \geqslant \delta_c) = \frac{1}{2} \dfrac{\exp[-u^2]}{u\sqrt{\pi}}
    \label{eq:ann:limitofPdelta}
,\end{equation}

\noindent and we can compute the asymptotic limit of $A_\text{p}$

\begin{equation}
    \lim_{\sigma \to 0} A_\text{p} = - \dfrac{\delta_c}{\sigma}
    \label{eq:ann:limitofAp}
,\end{equation}

\noindent which goes to $- \infty$ for $\delta_c > 0$. Then we can compute the limits of $A_\text{th}$ and $A_\text{turb}$ given Eq.\ref{eq:ann:athpturb}:

\begin{equation}
    \begin{cases}
        \lim_{\sigma \to 0} A_\text{th} = A_\text{p} \left[ 1 - \dfrac{d \ln T}{d \ln \rho} \bigg( 1 + \dfrac{\delta_c}{2} \bigg) \right] \\[15pt]
        \lim_{\sigma \to 0} A_\text{turb} = A_\text{p} \left[1 + \eta \dfrac{\delta_c}{2} \right]
    \end{cases}
    \label{eq:ann:limitsofAthAturb}
.\end{equation}
 
Using these asymptotic limits, the fragmentation rate $\phi(R)$ given Eq.\ref{eq:app:fragrate} can finally be written as

\begin{equation}
        \phi(R) \approx \xi + 3 - \frac{2 \bigg( 1 + \eta \dfrac{\delta_c}{2} \bigg)}{ 1 - \dfrac{d \ln T}{d \ln \rho} \bigg( 1 + \dfrac{\delta_c}{2} \bigg)}
.\end{equation}
            
\subsection{Supersonic limit}

Next we assume that $V_\text{rms} \gg C_s$ such that turbulent support dominates. In this limit, the Mach number $\mathcal{M}^2$ goes to $+\infty$ and so is the density fluctuations variance $\sigma^2$. We can compute in the supersonic limit the following asymptotic relations:

\begin{equation}
\begin{cases}
    \lim_{\sigma \to +\infty } u = \dfrac{\sigma}{2 \sqrt{2}} \\[15pt]
    \lim_{\sigma \to +\infty } v = - \dfrac{\sigma}{2 \sqrt{2}} \\[15pt]
    \lim_{\mathcal{M} \to +\infty } \hat{C_s}^2 = 0 \\[15pt]
    \lim_{\mathcal{M} \to +\infty } \hat{V}_\text{rms}^2 = 1 \\[15pt]
    \lim_{\mathcal{M} \to +\infty } \hat{\mathcal{M}}^2 = 1
\end{cases}
.\end{equation}

We can show that the asymptotic limit of the quantity $\dfrac{v \hat{\mathcal{M}}^2}{\sqrt{2} \sigma}$ involved within the coefficients $A_\text{th}$ and $A_\text{turb}$ is finite:

\begin{equation}
    \lim_{\sigma \to +\infty } \dfrac{v \hat{\mathcal{M}}^2}{\sqrt{2} \sigma} = -\dfrac{1}{4}
.\end{equation}

In this limit, the variable $u$ goes to $+\infty$ so the asymptotic limit of $A_\text{p}$ is given by Eq.\ref{eq:ann:limitofAp} and $A_\text{p}$ goes to 0 in the supersonic limit. Then we recompute the limits of $A_\text{th}$ and $A_\text{turb}$:

\begin{equation}
    \begin{cases}
        \lim_{\sigma \to 0} A_\text{th} = 0 \\[15pt]
        \lim_{\sigma \to 0} A_\text{turb} = 0
    \end{cases}
.\end{equation}
 
Using these asymptotic limits, the fragmentation rate $\phi(R)$ given Eq.\ref{eq:app:fragrate} can be simplified to

\begin{equation}
        \phi(R) \approx 3
.\end{equation}

\subsection{Case $\delta_c < 0$ in the subsonic limit}
\label{app:deltacnegative}

If $\delta_c < 0$, the density of a fragment is higher than its critical density and it does not need any turbulent fluctuation to collapse. On the contrary of the $\delta_c > 0$ case, the variable $u$ then goes to $-\infty$ in the subsonic limit. The error function $\text{erf}(u)$ is odd since $\text{erf}(-u) = -\text{erf}(u)$. In that case, the limit of Eq.\ref{eq:ann:limitofPdelta} becomes

\begin{equation}
    \lim_{u \to -\infty} \mathcal{P}(\delta \geqslant \delta_c) = \frac{1}{2} \bigg( 2 - \lim_{u \to -\infty} \dfrac{\exp[-u^2]}{u\sqrt{\pi}} \bigg) = 1
.\end{equation}

\noindent As $u \sim \sigma^{-1}$,

\begin{equation}
    \lim_{\sigma \to 0 } A_\text{p} = - \frac{\exp[-u^2]}{\sigma \sqrt{2\pi}} = 0
.\end{equation}
    
Hence, both $A_\text{th}$ and $A_\text{turb}$ goes to 0 as in the supersonic case and the fragmentation rate $\phi$ can be expressed in the subsonic limit when a fragment is produced in an unstable state:

\begin{equation}
    \phi(R) \approx 3 
,\end{equation}

\noindent which is independant of the mass transfer rate $\xi$.


\section{2D projection of 3D fragments}
\label{Appendix:comparisonNGC2264}

In Sect.\ref{Sec:Observational constraints on the fragmentation rate} we measure the two dimensional average fragmentation rate $\overline{\phi_{2D}}$ in {NGC~2264} molecular cloud from elliptical fragments extracted in \cite{thomasson2022}. In this appendix we aim to infer the three dimensional fragmentation rate $\overline{\phi_{3D}}$ and evaluate its uncertainty from any measurement of $\overline{\phi_{2D}}$. In order to calibrate the error caused by the projection of 3D ellipsoidal clumps to 2D elliptical clumps on the fragmentation rate, we employ a Monte-Carlo method to generate ellipsoids in a three-dimensional space and randomly place child ellipsoids inside them. 

We adopt similar conditions to the multi-scale structures extracted in \cite{thomasson2022} from elliptical fragments. First we consider child ellipsoids whose spatial extension is at least 75\% included inside their parent. Then we set a total of L~=~6 fragmentation levels separated by scaling ratios $r = 1.5$. In addition, ellipsoids on the same level are assumed to not intersect. 

The average number of children $\bar{n_l}$ to be placed in a parent is determined by Eq.\ref{eq:micro_n_connection}. Assuming a constant fragmentation rate $\phi$ between the first and last levels and that $\bar{n_l}$ is the same for all parents we have

\begin{equation}
    \bar{n_l} = r^\phi
.\end{equation}

A child is placed inside a parent taking into account that (i) a parent contains an integer number of fragments an (ii) on average a parent contains $r^\phi$ children for any fragmentation rate $\phi$. In order to calibrate a worst case scenario error we test the filling limits of a parent as a function of the fragmentation rate $\phi$ as we choose to select the number of children according to the following rule. Let $\floor{x}$ be the integer part of $x$. A parent can at most form a number $\floor{\bar{n_l}} + 1$ of children with probability $p = \bar{n_l} - \floor{\bar{n_l}}$; and at least form a number $\floor{\bar{n_l}}$ with probability $1 - p$. 

At any level each ellipsoid is constructed from the coordinates of its centroid $(x, y, z)$, its half-axes $(a, b, c)$ and its 3D orientation assumed to be isotropic. The half-axis $a$ of an ellipsoid of scale $R_l$ is selected randomly, according to a normal distribution $\mathcal{N}(\mu, \sigma)$ with mean $\mu = R_l$ and standard deviation $\sigma = 0.1R_l$, equivalent to 10\% of its characteristic scale to provide small fluctuations. The half-axis $b$ is defined as the half-small axis, selected from an aspect ratio sampled uniformly between 0.7 and 1, with mean 0.85. The half-axis $c$ is defined as the semi major axis, selected from an aspect ratio uniformly sampled between 1 and 1.4, with an average of 1.2. Since the typical scale $R_l$ of an ellipsoidal fragment is given by the product $(abc)^{1/3}$, we expect to generate, on average, objects of size

\begin{equation}
    (a \times 0.85a \times 1.2a)^{1/3} \approx  a
.\end{equation}

These are indeed objects of scale $R_l$. The centroid coordinates $(x, y, z)$ of the initial fragments at level $l = 0$ are selected uniformly within a cube of side $a_\text{cube}$. The coordinates of the centroid of the child fragments are selected uniformly within the rectangular cuboid enclosing their parent ellipsoid. 

From an initial population of {$N = 10^3$} ellipsoids at $l = 0$ we generate different child ellipsoids along the $L = 6$ fragmentation level. The 2D projection of these ellipsoids generates objects with elliptical geometry. Since the orientation of ellipsoids is isotropic all the projection planes are equivalent. In these conditions, the minimum aspect ratio that an ellipse projected in two dimensions can have in any plane is

\begin{equation}
    \frac{\text{min}(b)}{\text{max}(c)} = \frac{0.7}{1.4} \approx 0.5
,\end{equation}

\noindent similar to the fragments extracted in \cite{thomasson2022}. Due to projection effect, the spatial extensions of two resulting ellipses may partially overlap. If the two ellipses overlap perfectly, one is no longer identifiable from the other and they appear as a single object. The distinction between the ellipses is checked as a function of their centre-to-centre separation $d$. Suppose that the fragments extracted from an observed image have Gaussian profiles in two dimensions. Since a Gaussian profile extends to infinity, it is necessary to define a finite limit associated with the size of a fragment. This limit, which defines the border of the fragment, have to encapsulate almost all of its mass. As chosen by \cite{thomasson2022}, the characteristic scale $R$ of {NGC~2264} fragments is defined as the 2$\sigma$ extension of their Gaussian profile whose full width at half-maximum (FWHM) is associated with the resolution $\theta$ of the image from which they are extracted. The $2 \sigma$ extension of a Gaussian profile is related to its FWHM $\theta${, as}

\begin{equation}
    \theta = \sqrt{2 \ln(2)} 2 \sigma
.\end{equation}

Hence, if the minimum distance $d_\text{min}$ between two extracted fragments is smaller than the resolution $\theta$ of the image from which they originate, these fragments would not be distinguishable from each other (Fig. \ref{fig:ellipsoid_projection}). Therefore, it is possible to separate the two ellipses if their centre-to-centre distance satisfies

\begin{equation}
    d > d_\text{min} = \sqrt{2 \ln(2)} R 
    \label{app:eq:conditionfilter}
.\end{equation}

Using Eq.\ref{app:eq:conditionfilter} to filter the projected ellipeses we can measure the average two dimensional fragmentation rate $\overline{\phi_{2D}}$. We employ the same method as in Sect.\ref{Sec:Observational constraints on the fragmentation rate} by measuring the fractality coefficient \citep{thomasson2022}. Then we calibrate the fragmentation rate error by computing the relative variation between the fragmentation rate $\overline{\phi_{3D}}$ used to generate the ellipsoids and $\overline{\phi_{2D}}$ measured after projection. We estimate that the actual fragmentation rate $\overline{\phi_{3D}}$ that generated a population of ellipsoidal fragments associated with $\overline{\phi_{2D}} = 0.54 \pm 0.12$ corresponds to $\overline{\phi_{3D}} = 0.77 \pm 0.20$.

\begin{figure}
    \centering
    \includegraphics[width=7cm]{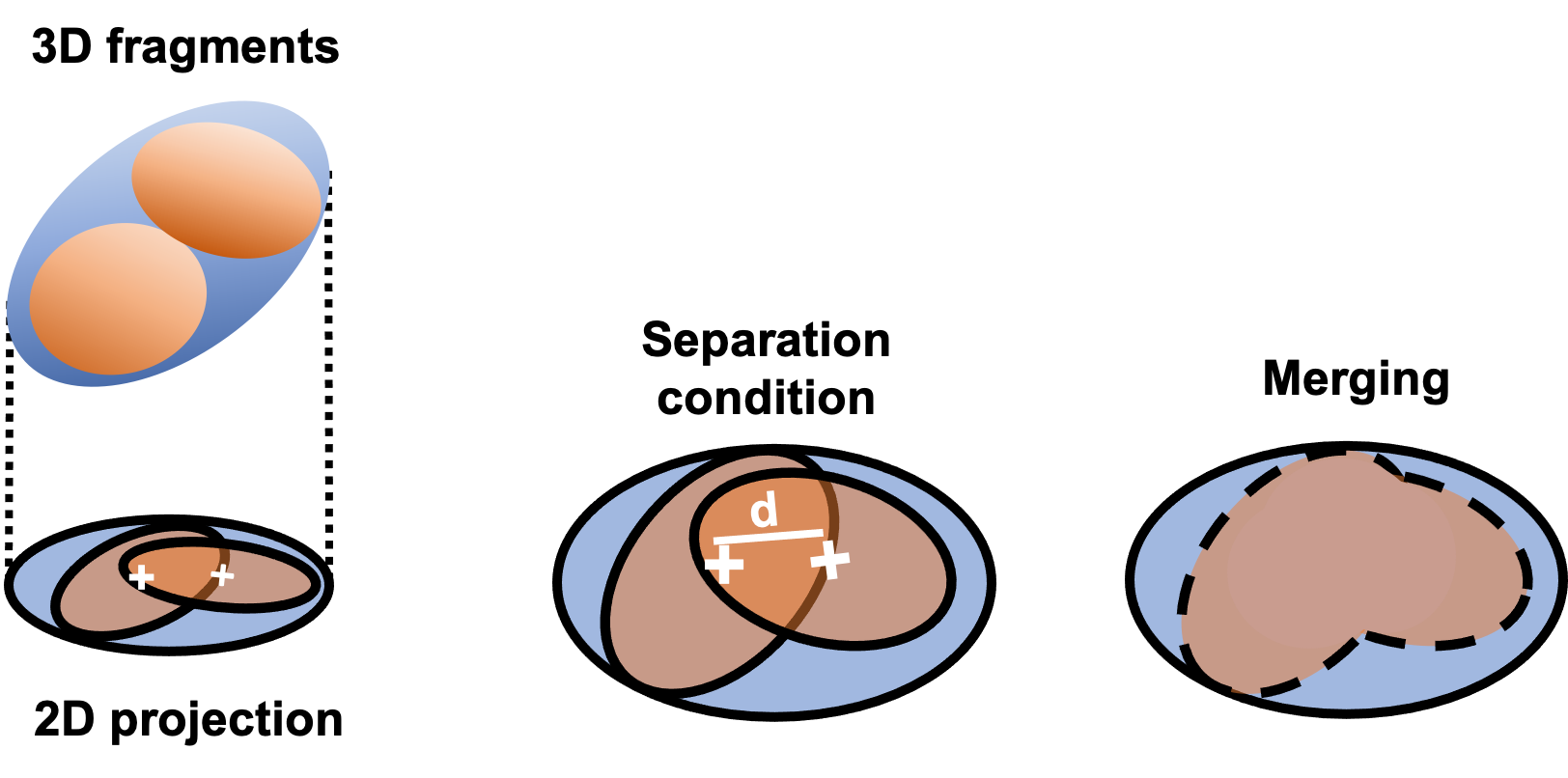}
    \caption{Projection of three-dimensonal ellipsoids into two dimensional ellipses. The parental (blue) object have fragmented into two child (orange) fragments. The merging condition of both child is regulated by their centre-to-centre distance $d$ and their 2D size according to Eq.\ref{app:eq:conditionfilter}. If they merge the number of children we may perceive in 2D is reduced, effectively reducing the two dimensional fragmentation rate one may measure.}
    \label{fig:ellipsoid_projection}
\end{figure}


\end{document}